\documentclass[12pt,reqno]{amsart}

\usepackage{amsmath}
\usepackage[foot]{amsaddr}
\usepackage{amssymb}

\usepackage{amsthm}

\usepackage{bm}

\usepackage{bbm}
\usepackage{graphicx}
\usepackage{xkeyval}

\usepackage{libertine}

\usepackage{libertinus}
\usepackage[T1]{fontenc}

\usepackage[nopatch=footnote]{microtype}
\usepackage{caption}
\usepackage{subcaption}

\usepackage[space]{grffile}
\usepackage{import}

\usepackage[dvipsnames]{xcolor}
\usepackage[hidelinks, colorlinks, urlcolor=BrickRed,linkcolor=black, citecolor=BrickRed]{hyperref}

\usepackage[nodisplayskipstretch]{setspace}
\usepackage{pdflscape}
\usepackage{float}
\usepackage[margin=1in]{geometry}
\usepackage[longnamesfirst,authoryear]{natbib}
\usepackage{multirow}
\usepackage{array}
\usepackage{dcolumn}
\usepackage{booktabs}
\usepackage{tablefootnote}
\usepackage{tikz}
\usetikzlibrary{positioning}

\usepackage{dsfont}

\let\hbar\relax

\newtheorem{lemma}{Lemma}
\newtheorem{proposition}{Proposition}
\newtheorem{corollary}{Corollary}

\newtheorem{definition}{Definition}

\usepackage{float}
\usepackage{placeins}
\usepackage{afterpage}

\begin{document}

\author{Marcos R. Fernandes$^{\dag}$}
\address[]{$\dag$ University of São Paulo and University of Rochester, Department of Economics, e-mail: \textsf{mrf.ross@gmail.com}.}
% \email{}
\dedicatory{This version: April 2026.}

%================================================================
\begin{abstract}
This paper studies how communication across experts prior to aggregation by a decision-maker affects the efficiency of forecast combination. When experts exchange information before reporting their forecasts, their signals become correlated through the communication network, altering aggregation efficiency even when forecasts are unbiased. The analysis introduces a statistic that characterizes how network structure shapes aggregation efficiency and shows that degree heterogeneity plays a central role. Among connected networks, regular networks attain the minimal level of aggregation distortion, while star networks generate the largest distortions within sparse connected structures. Random network benchmarks show that aggregation efficiency approaches the regular-network benchmark when expected degree either vanishes or becomes large as network size increases, whereas networks with constant expected degree generate intermediate distortions. These results provide a theoretical foundation for understanding how communication across experts affects forecast combination and establish a connection between the forecast combination literature and models of social learning in networks.
\\

\textit{JEL Classification:} C44, C53, D83, D85.

\vspace{2mm}
\textit{Keywords:} forecast combination, networks, experts, information aggregation, social learning.

\end{abstract}
%================================================================
\title{Combining Combined Forecasts: a Network Approach}

\thanks{I thank Luis Alvarez, Patrick Allmis, Lawrence Blume, H\"ulya Eraslan, Ben Golub, Hari Govindan, Matthew O. Jackson, Ming Li, Mihai Manea, Yusuf Masatlioglu, Marcos Nakaguma, Marciano Siniscalchi, William Thomson, Asher Wolinsky, and Huan Xie for very helpful comments and suggestions. I also thank participants at the 23rd Annual SAET Conference, the 9th Annual Network Science and Economics Conference, the 35th Stony Brook International Conference on Game Theory, the 2024 LACEA–LAMES Meeting, the 2024 Society for Economic Dynamics Winter Meeting, the Fall 2025 Midwest Theory Conference, and the Theory Workshops at the University of Rochester and UC Davis. This project has received funding from the University of São Paulo (USP) through the Incoming Assistant Professor Fellowship (Bolsa Novos Doutores) and from Fundação Instituto de Pesquisas Econômicas (FIPE). Acknowledgment: Both \textcolor{blue}{\textsf{Refine.ink}} and \textcolor{blue}{\textsf{Coarse.ink}} were used to check the paper for both consistency and clarity}

\maketitle

% ===========================================================
\section{Introduction}
\label{sec:intro}
% ===========================================================
The practice of combining forecasts is widely used to improve prediction accuracy by aggregating information from multiple sources. Even simple aggregation rules, such as equal-weight averaging, often perform comparably to more complex methods in empirical applications. Since the seminal contribution of \cite{bates1969combination}, a large literature has examined the statistical and decision-theoretic properties of forecast combinations under different assumptions about forecast precision, dependence, and loss functions \citep{elliott2004optimal}. An important challenge in this literature concerns the role of interaction among experts prior to aggregation. When experts exchange information before reporting their forecasts, their assessments may become correlated through communication, altering the informational content of the forecasts received by the decision-maker.

This paper studies how communication across experts prior to aggregation affects the efficiency of forecast combination. In many environments experts do not report their assessments independently. Instead, they exchange information with peers before communicating their final forecasts to a decision-maker who may not observe the structure of these interactions. Such communication modifies the dependence structure of forecasts and therefore affects the precision of aggregated information even when experts are unbiased and equally informative.

The relevance of this question arises in several contexts in which expert assessments are formed through interaction. In health-care planning, experts may exchange information when evaluating the adoption of new technologies or determining staffing requirements for medical centers. In legal settings, expert witnesses may consult with other specialists before submitting their final reports to the court. In macroeconomic forecasting, professional forecasters may exchange views before reporting expectations to policy institutions. More broadly, similar mechanisms arise in environments where individuals form evaluations after consulting peers prior to submitting ratings or recommendations.

To illustrate the mechanism studied in the paper, consider the introductory example shown in Figure~\ref{fig:intro-ex}. Suppose three experts provide point forecasts to a decision-maker who aggregates them using their simple average, corresponding to the optimal linear aggregation rule under equal precision and equicorrelation across forecasts. The white nodes represent forecasts reported without prior communication, while the gray nodes represent forecasts after experts update their assessments using local averages within their communication neighborhoods. In the first row of each panel the communication links are shown as dashed edges to indicate that they represent potential interaction paths that become active only when experts update their forecasts prior to reporting them to the decision-maker.

If experts do not communicate, the decision-maker receives the original forecasts. If they communicate locally before reporting their assessments, each expert first updates its forecast using the information available through the network and then reports the updated value. As illustrated in the three cases shown in Figure~\ref{fig:intro-ex}, the differences arise because communication changes the forecasts reported by experts prior to aggregation, so that the location of the initial forecasts within the network affects how information enters the decision-maker's final average.

\medskip
\begin{figure}[H]
        \centering
        \begin{subfigure}[b]{0.32\textwidth}   
            \centering
            \begin{tikzpicture}
  [scale=1.5,auto=left,every node/.style={circle,fill=white,draw=black,line width=0.8pt,scale=0.9}]
  \node (n1) at (1,2)  {1};
  \node (n2) at (2,2)  {5};
  \node (n3) at (3,2)  {3};
  
      \foreach \from/\to in {n1/n2, n2/n3}
    \draw[dashed,line width=0.22mm] (\from) -- (\to);
\end{tikzpicture}
            \vspace{4 mm}\\
            \centering
            \begin{tikzpicture}
  [scale=1.5,auto=left,every node/.style={circle,fill=gray!20,draw=black,line width=0.8pt,scale=0.9}]
  \node (n1) at (1,2)  {3};
  \node (n2) at (2,2)  {3};
  \node (n3) at (3,2)  {4};
  
      \foreach \from/\to in {n1/n2, n2/n3}
    \draw[line width=0.22mm] (\from) -- (\to);
\end{tikzpicture}
            \vspace{3 mm}
            \caption{Case 1: communication alters the aggregate forecast from $(1+5+3)/3 = 3$ to $(3+3+4)/3 = 10/3$}   
            \label{fig:intro-ex-c1}
        \end{subfigure}
        \hfill    
        \begin{subfigure}[b]{0.32\textwidth}
            \centering
            \begin{tikzpicture}
  [scale=1.5,auto=left,every node/.style={circle,fill=white,draw=black,line width=0.8pt,scale=0.9}]
  \node (n1) at (1,2)  {1};
  \node (n2) at (2,2)  {3};
  \node (n3) at (3,2)  {5};
  
      \foreach \from/\to in {n1/n2, n2/n3}
    \draw[dashed,line width=0.22mm] (\from) -- (\to);
\end{tikzpicture}
            \vspace{4 mm}\\
            \centering
            \begin{tikzpicture}
  [scale=1.5,auto=left,every node/.style={circle,fill=gray!20,draw=black,line width=0.8pt,scale=0.9}]
  \node (n1) at (1,2)  {2};
  \node (n2) at (2,2)  {3};
  \node (n3) at (3,2)  {4};
  
      \foreach \from/\to in {n1/n2, n2/n3}
    \draw[line width=0.22mm] (\from) -- (\to);
\end{tikzpicture}
            \vspace{3 mm}
            \caption{Case 2: communication leaves the aggregate forecast unchanged at $3$} 
            \label{fig:intro-ex-c2}
        \end{subfigure}
        \hfill
        \begin{subfigure}[b]{0.32\textwidth}  
            \centering
            \begin{tikzpicture}
  [scale=1.5,auto=left,every node/.style={circle,fill=white,draw=black,line width=0.8pt,scale=0.9}]
  \node (n1) at (1,2)  {3};
  \node (n2) at (2,2)  {1};
  \node (n3) at (3,2)  {5};
  
      \foreach \from/\to in {n1/n2, n2/n3}
    \draw[dashed,line width=0.22mm] (\from) -- (\to);
\end{tikzpicture}
            \vspace{4 mm}\\
            \centering
            \begin{tikzpicture}
  [scale=1.5,auto=left,every node/.style={circle,fill=gray!20,draw=black,line width=0.8pt,scale=0.9}]
  \node (n1) at (1,2)  {2};
  \node (n2) at (2,2)  {3};
  \node (n3) at (3,2)  {3};
  
      \foreach \from/\to in {n1/n2, n2/n3}
    \draw[line width=0.22mm] (\from) -- (\to);
\end{tikzpicture}
            \vspace{3 mm}
            \caption{Case 3: communication alters the aggregate forecast from $(1+5+3)/3 = 3$ to $(2+3+3)/3 = 8/3$}    
            \label{fig:intro-ex-c3}
        \end{subfigure}
\caption{Example of forecast aggregation with and without prior communication across experts \\ \footnotesize{White nodes: forecasts reported without prior communication (links inactive). \\ Gray nodes: forecasts reported after local communication across connected experts.}} 
\label{fig:intro-ex}
\end{figure}

This example illustrates that both the realization of forecasts and the structure of communication across experts affect the information ultimately available to the decision-maker. The objective of the paper is to characterize these effects in a general framework in which forecasts are stochastic and communication occurs through an arbitrary network structure.

The analysis introduces a statistic, denoted $Q(G(n,A))$, that summarizes how communication across experts shapes aggregation efficiency when forecasts become correlated through local interaction. This statistic depends on weighted counts of indirect information paths in the network and provides a tractable measure for comparing aggregation performance across network structures.

The results show that aggregation efficiency depends primarily on the degree distribution of the network rather than on its density alone. Within the class of connected networks, $d$-regular networks attain the minimal value of $Q(G(n,A))$, while star networks generate the largest values within sparse connected structures. These extremal structures clarify how degree heterogeneity distorts the effective contribution of experts’ signals to the data-generating process observed by the decision-maker. Regular networks ensure that information enters the communication-induced data-generating process symmetrically across experts, whereas star networks concentrate indirect influence on a central expert and therefore generate the largest aggregation distortions within this class.

Random network benchmarks complement these deterministic results. In the Poisson random graph model, aggregation efficiency converges to the regular-network benchmark when the expected degree either vanishes or diverges with network size, while persistent degree heterogeneity in the constant-degree regime generates a strictly larger asymptotic distortion factor. Preferential attachment networks exhibit slower convergence due to the presence of hubs and therefore display intermediate behavior between regular and star networks. Taken together, these results establish a hierarchy of aggregation efficiency across network classes indexed by their degree heterogeneity.

These findings contribute to the forecast combination literature by showing that communication across experts prior to aggregation can substantially affect the informational content of combined forecasts even when experts are unbiased. More broadly, the analysis provides a theoretical foundation for understanding how communication structure shapes the quality of expert-based decision-making and establishes a connection between forecast combination methods and models of social learning in networks. Taken together, the results show that aggregation efficiency in large networks is governed primarily by degree heterogeneity rather than by network density itself.

%============================================================
\section{Literature review and contribution}
\label{sec:litrev}

The objective of combining forecasts is to improve predictive accuracy by aggregating information from different sources and reducing data, parameter, and model uncertainty. The seminal method introduced by \cite{bates1969combination}, and extended by \cite{newbold1974experience} to combinations involving more than two forecasts, determines optimal weights by minimizing the variance of the combined forecast error. These weights depend on the variance–covariance structure of forecast errors, which is typically unknown in practice or costly to estimate. For this reason, the simple equal-weight average has emerged as a popular and robust combination rule (\cite{makridakis1982accuracy, clemen1989combining, timmermann2006forecast, hsiao2014there, makridakis2020m4}).

The literature has also examined alternative aggregation rules, including the median, mode, and trimmed or winsorized means (\cite{genre2013combining, jose2014trimmed, grushka2017ensembles}). A substantial literature has developed around the combination of individual forecasts, with comprehensive reviews including \cite{genest1986combining, granger1989invited, clemen1989combining, jacobs1995methods, timmermann2006forecast, mancuso2013review, gneiting2014probabilistic, wallis2014combining, wang2023forecast}.

Forecasts combination has also found applications in a wide range of fields, including energy \citep{xie2016gefcom2014, nielsen2007optimal}, retail \citep{ma2021retail}, tourism \citep{andrawis2011combination}, inflation forecasting \citep{mitchell2005evaluating}, epidemiology \citep{ray2023comparing}, health \citep{lipscomb1998combining}, weather \citep{gneiting2005weather}, environmental studies \citep{westerlund2014application}, and economics more broadly \citep{aastveit2018evolution}.

An important aspect of combining information from multiple sources concerns the potential presence of dependence across forecasts. Early contributions studied Bayesian approaches for updating forecast-combination weights in the presence of such dependence. Assuming that the vector of forecast errors follows a normal distribution, \cite{winkler1981combining} and \cite{clemen1985limits, clemen1986combining} developed Bayesian procedures based on conjugate priors for the variance–covariance matrix. Subsequent work by \cite{diebold1990use} incorporated the standard normal–gamma conjugate prior in a regression-based combination framework. More recently, attention has turned to environments in which the correlation structure across information sources is itself uncertain, and to aggregation procedures that remain robust under such uncertainty \citep{arieli2018robust, levy2021maximum, levy2022combining}. This literature highlights that the effectiveness of forecast combination depends not only on the number of forecasts available but also on the dependence structure linking them.

Dependence across forecasts may also arise because \textit{experts exchange information prior to reporting their forecasts to a decision-maker}, so that the forecasts ultimately combined may already incorporate locally aggregated information. Such environments are common in practice, for example when professional forecasters discuss projections before reporting to policy institutions, or when experts exchange views before issuing formal recommendations. 

Interaction among experts can be naturally represented using networks, which determine how information flows across experts before aggregation takes place. A large literature studies how beliefs evolve through interaction in networks and how network structure affects the aggregation of dispersed information.

An early branch of the social-learning literature studied herd behavior in sequential environments, where agents take actions after observing predecessors’ actions and private signals \citep{banerjee1992simple, bikhchandani1992theory, smith2000pathological}. More recent contributions extend these classical herding frameworks to network environments in which agents observe only subsets of past actions \citep{ccelen2004observational, eyster2010naive, acemoglu2011bayesian, lobel2015information}. While sequential learning models provide important insights into informational cascades and observational inference, they do not fully capture environments in which agents interact simultaneously and influence one another through repeated communication prior to reporting their assessments.

A complementary strand of the literature studies belief updating under repeated interaction in networks. Early contributions analyzed how network structure and initial opinions shape consensus formation and its speed \citep{degroot1974reaching, chatterjee1977towards, chatterjee1987combining, friedkin1999influence, demarzo2003persuasion, golub2010naive}, as well as conditions under which consensus fails to emerge \citep{hegselmann2002opinion, yildiz2013binary}. Comprehensive overviews of this literature are provided by \cite{golub2017learning} and \cite{grabisch2020survey}. More recent research incorporates Bayesian updating, behavioral biases, conformism, selective information sharing, and misinformation into network-based learning environments \citep{jadbabaie2012non, mueller2013general, mossel2015strategic, buechel2015opinion, li2020locally, azzimonti2023social, fernandes2023confirmation, buechel2023misinformation}.

This study links these two strands of research by analyzing forecast combination in environments where \textit{dependence across forecasts emerges through local interaction among experts} prior to decision-maker aggregation. In contrast to standard forecast combination frameworks in which forecasts are treated as independent summaries of private information, we consider a setting in which experts first revise their forecasts locally through communication with their neighbors and are subsequently aggregated by an \textit{external} decision-maker who is \textit{unaware} of the underlying communication structure. Under homogeneous signal precision and common correlation across experts, we show that such interaction does not introduce bias in the combined forecast but generates topology-dependent differences in efficiency summarized by a simple graph functional. In particular, highly centralized structures lead to inefficient aggregation, whereas more evenly connected networks approach the efficient benchmark as the number of experts increases. These results provide a tractable characterization of how network topology shapes the efficiency of forecast combination when forecasts already incorporate locally aggregated information.

% ====================================================
\section{Model}
\label{sec:model}
% ====================================================

The primitives of the model consist of two key elements: an information structure and a network structure.  
The information structure describes the forecasts produced by experts about an unknown quantity of interest, while the network structure determines how these forecasts are shared prior to communication with the decision-maker.

\vspace{2mm}
\paragraph{\textit{Experts’ forecasts and forecast errors}.}

Let $\theta \in \mathbb{R}$ denote the quantity to be forecast, representing either a future realization or an unknown parameter.  
A group of $n$ experts, indexed by $i=1,\ldots,n$, provide forecasts about $\theta$. Let $N=\{1,\ldots,n\}$ denote the set of experts. Each expert $i$ issues a point forecast $x_i$, which we interpret as the mean of expert $i$’s subjective predictive distribution under quadratic loss. Forecast errors are defined as $\varepsilon_i = x_i - \theta$.

Following \cite{winkler1981combining} and \cite{clemen1985limits, clemen1986combining}, the vector of forecast errors $\bm{\varepsilon}=(\varepsilon_1,\ldots,\varepsilon_n)'$ is assumed to be jointly normally distributed with mean zero and covariance matrix $\bm{\Sigma}$. Equivalently, $\bm{x}\mid \theta \sim \mathcal{N}(\theta\bm{1},\bm{\Sigma})$. The diagonal element $\Sigma_{ii}=\sigma_i^2$ represents the variance of expert $i$’s forecast error and measures the precision of expert $i$’s forecast, $\tau_i=\sigma_i^{-2}$. The off-diagonal elements of $\bm{\Sigma}$ capture statistical dependence across forecasts, reflecting common information sources or similarities in forecasting methods. 

In what follows, we focus on a symmetric benchmark environment with equal forecast error variances and common pairwise correlation across experts. In particular, we assume $\Sigma_{ii}=\sigma^2$ and $\Sigma_{ij}=\rho\sigma^2$ for all $i\neq j$, where $\rho\in[0,1)$ denotes the correlation parameter. This equicorrelated covariance structure serves as the maintained benchmark throughout the paper and allows aggregation efficiency to be characterized in closed form.

\vspace{2mm}
\paragraph{\textit{Experts’ network}.}

Before communicating with the decision-maker, experts are allowed to observe the forecasts of a subset of other experts. These interactions are represented by an undirected network $G(N,A)$, where $A$ is a binary $n\times n$ adjacency matrix and $A_{ij}=1$ indicates that expert $i$ observes expert $j$’s forecast. Throughout the paper we assume that each expert observes their own forecast, so that $A_{ii}=1$ for all $i\in N$, and that the network is connected. Accordingly, the neighborhood of expert $i$ is defined as $N_i=\{j\in N:\,A_{ij}=1\}$ with $i\in N_i$, and the degree of expert $i$, $d_i=\sum_{j\in N}A_{ij}$, represents the number of forecasts observed by expert $i$ prior to communicating with the decision-maker, including their own. Connectivity together with self-observation implies that each expert observes at least one other expert’s forecast, so that $d_i\ge 2$ for all $i\in N$.

Communication occurs in a single round: each expert simultaneously observes the original forecasts of agents in their neighborhood and updates their assessment of $\theta$ based on this information prior to reporting to the decision-maker.

\vspace{2mm}
\paragraph{\textit{Posterior assessments}.}

The decision-maker and all experts are assumed to hold diffuse improper priors over $\theta$, so that posterior assessments depend entirely on the observed forecasts and their joint error structure. Under the normality assumption above, observing any subset $S\subseteq N$ of forecasts implies
\[
\theta\mid \bm{x}_S
\sim
\mathcal{N}\!\left(
\frac{\bm{1}_S'\bm{\Sigma}_S^{-1}\bm{x}_S}
{\bm{1}_S'\bm{\Sigma}_S^{-1}\bm{1}_S},
\;
(\bm{1}_S'\bm{\Sigma}_S^{-1}\bm{1}_S)^{-1}
\right).
\]

Accordingly, expert $i$’s posterior assessment after observing the forecasts of neighbors in $N_i$, including its own forecast, under the maintained diffuse prior over $\theta$, is
\[
\tilde{\theta}_i(\bm{x})
=
\frac{\bm{1}_{N_i}'\bm{\Sigma}_{N_i}^{-1}\bm{x}_{N_i}}
{\bm{1}_{N_i}'\bm{\Sigma}_{N_i}^{-1}\bm{1}_{N_i}},
\]
with posterior variance
\[
\tilde{\sigma}_i^2
=
\left(
\bm{1}_{N_i}'\bm{\Sigma}_{N_i}^{-1}\bm{1}_{N_i}
\right)^{-1}.
\]

By contrast, a decision-maker who directly observes the \textit{full vector} of forecasts $\bm{x}$ forms the benchmark posterior assessment
\[
\tilde{\theta}_0(\bm{x})
=
\frac{\bm{1}'\bm{\Sigma}^{-1}\bm{x}}
{\bm{1}'\bm{\Sigma}^{-1}\bm{1}},
\]
with posterior variance
\[
\tilde{\sigma}_0^2
=
\left(
\bm{1}'\bm{\Sigma}^{-1}\bm{1}
\right)^{-1}.
\]

% =================================================
\section{Main Results}
\label{sec:main-res}
% ==================================================

This section characterizes the statistical properties of forecast aggregation when experts exchange information through a communication network prior to reporting their forecasts to the decision-maker. The results identify how network structure affects aggregation precision through a single statistic that summarizes indirect information flows across experts. Proofs of all main results are presented in Appendix~\ref{append:proofs-main-res}, whereas proofs of auxiliary lemmas are presented in Appendix~\ref{append:proof-aux-lemmas}.

% ==========================================================
\subsection{Benchmark Aggregation and the CCF Estimator}
\label{subsec:bench-agg-ccf}
% ==========================================================

As a benchmark, we first extend the aggregation result of \cite{winkler1981combining} to a network environment in which each expert observes only a subset of forecasts.

% ---------------------------------------------------------
\begin{proposition}
\label{prop:equiv-rules}
For a decision-maker who observes the full forecast vector $\bm{x}$, the posterior mean and variance of $\theta$ are given by
\begin{equation}
\label{eq:mean-var-DM}
\Tilde{\theta}_0(\bm{x}) = \frac{1}{n}\sum_{i=1}^{n}x_i,
\quad
\Tilde{\sigma}_{0}^{2} = \frac{\sigma^2}{n}\left[1+(n-1)\rho\right],
\end{equation}
whereas for any expert $i\in N$ who observes only the subset of forecasts $\bm{x}_{N_i}$ of size $d_i$, the posterior mean and variance are given by
\begin{equation}
\label{eq:mean-var-nodes}
\Tilde{\theta}_i(\bm{x}) = \frac{1}{d_i}\sum_{j\in N_i}x_j,
\quad
\Tilde{\sigma}_i^2 = \frac{\sigma^2}{d_i}\left[1+(d_i-1)\rho\right].
\end{equation}
\end{proposition}
% -----------------------------------------------------------

Equation~\eqref{eq:mean-var-DM} reproduces the aggregation rule in \cite{winkler1981combining}. In the equicorrelated equal-precision environment considered here, simple averaging coincides with the optimal linear aggregation rule and the posterior variance of the decision-maker is bounded below by $\sigma^2\rho$ as the number of experts increases. Positive correlation therefore limits the information gain from consulting additional forecasts. Equation~\eqref{eq:mean-var-nodes} describes the same updating problem from the perspective of an expert embedded in a network. In this case posterior precision depends on the number of forecasts observed locally, and therefore on network structure through the degree $d_i$.

When $n>d_i$, the decision-maker observes strictly more information than any individual expert because she aggregates the original forecasts directly. 

We now consider the case in which experts communicate before reporting to the decision-maker. Instead of observing the original forecast vector $\bm{x}$, the decision-maker observes the vector of experts’ posterior means. This leads to the combining-combined-forecasts estimator defined below.

% ---------------------------------------------------------
\begin{definition}[Combining Combined Forecasts (CCF) estimator]
\label{def:CCF}
The decision-maker is said to be \emph{combining combined forecasts} when, unaware of the network structure $G(n,A)$, she updates her belief using the vector $\Tilde{\bm{\theta}}(\bm{x}) = (\Tilde{\theta}_1(\bm{x}), \ldots, \Tilde{\theta}_n(\bm{x}))'$ of experts’ posterior means—each based on local information. Her posterior mean is then given by
\begin{equation}
\label{eq:ccf-agg-linform}
\Tilde{\theta}_0\left(\Tilde{\bm{\theta}}(\bm{x})\right) = \left(\bm{1}' \bm{1}\right)^{-1}\bm{1}' D^{-1}A\bm{x},    
\end{equation}
where $\Tilde{\theta}_0(\cdot)$ denotes the Bayesian updating rule applied to the vector of posterior means, with $\Tilde{\bm{\theta}}(\bm{x}) = D^{-1}A\bm{x}$.
\end{definition}
% ---------------------------------------------------------

Because the decision-maker does not observe the communication network, she treats the reported forecasts as primitive signals and applies the aggregation rule that is optimal under the maintained equal-variance equicorrelated covariance structure. After communication across experts, however, the dependence structure of the reported forecasts is altered by the network and therefore differs from the benchmark structure assumed by the decision-maker. As a result, the CCF estimator generally does not coincide with the Bayesian posterior conditional on the true communication-induced dependence structure.

If the communication network were observed, the decision-maker could in principle account for the transformation relating original forecasts to reported forecasts. When this transformation is invertible, the original forecast vector could be recovered prior to aggregation. The analysis therefore focuses on environments in which interaction patterns across experts are not observed by the decision-maker.

% ==========================================================
\subsection{Aggregation Efficiency and the Network Statistic \texorpdfstring{$Q(G(n,A))$}{Q(G(n,A))}}
\label{subsec:agg-eff-stat-Q}
% ==========================================================

We next characterize the statistical properties of the CCF estimator. In contrast to the benchmark case, aggregation efficiency now depends on the network through the structure of information exchange among experts.

% ---------------------------------------------------------
\begin{proposition}
\label{prop:mean-var-ccf}
Let $G(n, A)$ be a connected network with $n \geq 3$ nodes. Then,
\begin{equation}
\label{eq:exp-mean-CCF}
\mathbb{E}\left(\Tilde{\theta}_0\left(\Tilde{\bm{\theta}}(\bm{x})\right) \mid G(n, A)\right) = \theta,
\end{equation} and
\begin{equation}
\label{eq:exp-var-CCF}
    \text{Var}\left(\Tilde{\theta}_0\left(\Tilde{\bm{\theta}}(\bm{x})\right) \mid G(n, A)\right) = \sigma^2 \left[\rho + (1 - \rho) Q(G(n,A))\right],
\end{equation}
where
\begin{equation}
    Q(G(n,A)) = \frac{1}{n^2} \sum_{i=1}^{n} \sum_{j=1}^{n} \frac{(A^2)_{ij}}{d_i d_j}.
\end{equation}
\end{proposition}
% ---------------------------------------------------------

The statistic $Q(G(n,A))$ therefore provides a sufficient network summary for comparing aggregation efficiency across network structures. Comparing \eqref{eq:mean-var-DM} with \eqref{eq:exp-mean-CCF} and \eqref{eq:exp-var-CCF} shows that network structure affects aggregation efficiency but not unbiasedness. The estimator remains centered at $\theta$, while its variance depends on the topology through the statistic $Q(G(n,A))$. 

Comparing variances before and after communication shows that aggregation based on forecasts reported after interaction is less efficient than the benchmark aggregation of original forecasts precisely when $Q(G(n,A))>\frac{1}{n}$.

To understand why $Q(G(n,A))$ captures the role of network topology, note that it aggregates weighted counts of walks of length two, where each walk is scaled by the degrees of its endpoints. Since $(A^2)_{ij}$ counts the number of intermediate experts through whom information can travel from expert $i$ to expert $j$ in two steps, the statistic measures how strongly experts are indirectly connected through common neighbors. The normalization by degrees assigns greater weight to connections involving low-degree experts, thereby reflecting the interaction between indirect information flows and degree heterogeneity in the network.

In particular, when $\rho=1$, forecasts are perfectly correlated and the posterior variance equals $\sigma^2$, so the communication \emph{network plays no role} for aggregation efficiency. When $\rho=0$, aggregation efficiency is entirely determined by $Q(G(n,A))$.

The next results characterize how this statistic varies across network structures and identify the topologies that attain its extremal values.

% ----------------------------------------------------------
\subsection{Bounds and Extremal Network Structures}
\label{subsec:extr-nets}
% ----------------------------------------------------------

The previous results show that aggregation efficiency depends on the network only through the statistic $Q(G(n,A))$. We now characterize how this statistic varies across network structures and identify topologies that attain its extremal values. The following proposition establishes bounds for $Q(G(n,A))$ and provides an asymptotic benchmark for the class of sparse connected structures.

Throughout the paper, we refer to sparse connected structures as connected networks whose underlying simple graph contains the minimal number of links required for connectivity (i.e., tree structures).

\begin{proposition}
\label{prop:bounds-Q-tree}
Let $G(n,A)$ be a connected network. Then
\[
Q(G(n,A)) \ge \frac{1}{n}.
\]
Moreover, if $G(n,A)$ is a sparse connected structure, then
\[
Q(G(n,A)) \le \frac{1}{4} + \frac{1}{n} - \frac{1}{n^2}.
\]
Consequently,
\[
0 \le \liminf_{n\to\infty} Q(G(n,A))
\quad\text{and}\quad
\limsup_{n\to\infty} Q(G(n,A)) \le \frac{1}{4}.
\]

\end{proposition}

The lower bound follows from the normalization implied by the structure of indirect information flows across experts and holds for \emph{all} connected networks. The upper bound is obtained using an edge-count argument applied to sparse connected structures, which contain the minimal number of links required for connectivity. Since this bound is increasing and convex in the number of edges, allowing denser connected graphs only weakens the implication of the inequality. A tight upper bound for arbitrary connected networks would require exploiting finer topological features than the number of edges alone. Sparse connected structures therefore provide a natural benchmark for assessing the largest aggregation distortions within this class.

The statistic $Q(G(n,A))$ depends on weighted counts of walks of length two, where each walk is scaled by the degrees of its endpoints. Networks that generate many such walks while concentrating connectivity in a small number of high-degree nodes therefore tend to attain large values of $Q(G(n,A))$, particularly within the class of sparse connected structures. The star network provides a canonical example of this configuration.

% ---------------------------------------------------------
\begin{proposition}
\label{prop:ineff-star}
Let $G(n,A)$ be a star network on $n \geq 3$ nodes. Then,
\[
Q(G(n,A)) = \frac{1}{n^2}\left[\frac{1}{n} + \frac{n-1}{2} + \frac{2(n-1)}{n} + \frac{(n-1)(n-2)}{4}\right].
\] 
Consequently,
\[
\lim_{n \to \infty} Q(G(n,A)) = \frac{1}{4},
\]
which corresponds to the largest asymptotic value attained by $Q(G(n,A))$ within sparse connected networks.
\end{proposition}
% ---------------------------------------------------------

The four terms in the expression correspond respectively to contributions from walks starting and ending at the hub, walks starting and ending at peripheral nodes, walks between the hub and peripheral nodes, and walks between pairs of peripheral nodes. As the network grows, only the contribution of walks between peripheral nodes remains asymptotically relevant. These walks dominate because they are both numerous and involve endpoints with low degrees, which receive larger weights in the definition of $Q(G(n,A))$. Consequently, the star network maximizes the asymptotic value of $Q(G(n,A))$ within sparse connected network structures.

To illustrate the quantitative magnitude of the efficiency loss, Table~\ref{tab:rel-eff-loss-star} reports the relative increase in the variance of the CCF estimator under star networks compared with the full-information benchmark (i.e., Equation \eqref{eq:mean-var-DM}) for representative values of $n$ and $\rho$. The results show that aggregation distortions can be substantial when correlation across forecasts is moderate and the number of experts is large. For example, when $\rho=0.2$, the posterior variance under the CCF estimator exceeds the benchmark variance by more than 90\% for $n=100$. As $\rho$ increases, the distortion becomes smaller because correlation already limits the marginal informational value of additional forecasts. These calculations illustrate that the effect of network-induced dependence on aggregation efficiency is quantitatively meaningful in environments with weakly correlated forecasts.\footnote{We define efficiency loss as the proportional increase in the variance generated by network-induced dependence across reported forecasts relative to the benchmark.}

\begin{table}[H]
\centering
\small
\begin{tabular}{ccc}
\hline
$n$ & $\rho$ & Efficiency loss \\
\hline
10 & 0.2 & 41.1\% \\
50 & 0.2 & 83.6\% \\
100 & 0.2 & 91.4\% \\
10 & 0.5 & 13.1\% \\
50 & 0.5 & 22.1\% \\
100 & 0.5 & 23.5\% \\
10 & 0.8 & 3.5\% \\
50 & 0.8 & 5.6\% \\
100 & 0.8 & 5.9\% \\
\hline
\end{tabular}
\caption{Relative efficiency loss under star networks}
\label{tab:rel-eff-loss-star}
\end{table}

\begin{figure}[H]
\centering

% --- Left: TikZ star network ---
\begin{minipage}[c]{0.45\textwidth}
\centering
\begin{tikzpicture}[scale=1.2,
  every node/.style={circle, draw, fill=white, minimum size=6mm},
  every path/.style={thick},
  font=\sffamily]

  % Center node
  \node (1) at (0,0) {1};

  % Leaf nodes
  \node (2) at (150:2)   {2};
  \node (3) at (90:2)    {3};
  \node (4) at (30:2)    {4};
  \node (5) at (-45:2)   {5};
  \node (n) at (225:2)   {$n$};

  % Rounded dots
  \node[draw=none, fill=none] at (260:2) {$\textbf{.}$};
  \node[draw=none, fill=none] at (270:2.025) {$\textbf{.}$};
  \node[draw=none, fill=none] at (280:2) {$\textbf{.}$};

  % Edges
  \foreach \i in {2,3,4,5,n} {
    \draw (1) -- (\i);
  }

\end{tikzpicture}
\caption*{\footnotesize{(a) Star network with $n$ nodes}}
\end{minipage}
% --- Right: PDF plot ---
\begin{minipage}[c]{0.52\textwidth}
\centering
\includegraphics[width=\linewidth]{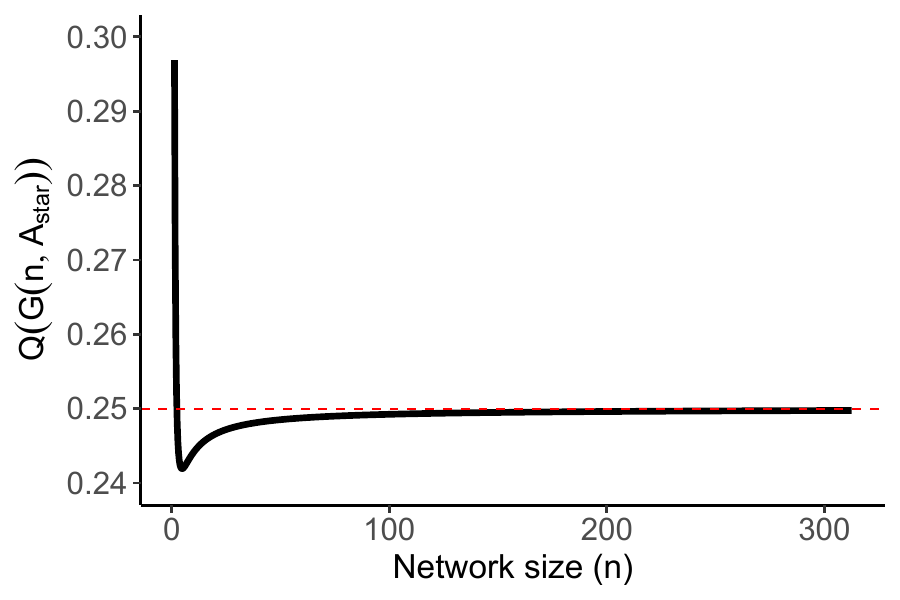}
\caption*{\footnotesize{(b) Non-monotonicity of $Q(G(n, A_{\text{star}}))$}}
\end{minipage}

\caption{Star network and corresponding statistic $Q(G(n,A))$.}
\label{fig:star_Q}
\end{figure}

Although the star network maximizes $Q(G(n,A))$ asymptotically within sparse connected structures, its finite-sample behavior is non-monotonic in the number of experts, as shown in panel (b) of Figure~\ref{fig:star_Q}. This implies that aggregation precision may be optimized at an interior network size. Corollary~\ref{cor:star-15-min} characterizes how communication across experts affects the variance of the CCF estimator relative to the benchmark posterior variance obtained when the decision-maker directly observes the original forecasts.

% ---------------------------------------------------------
\begin{corollary}
\label{cor:star-15-min}
In a star network, the variance of the CCF estimator is minimized when the number of experts consulted is $n = 15$.
\end{corollary}
% ---------------------------------------------------------

The intuition behind this result is that increasing the number of experts initially improves aggregation efficiency, since the decision-maker gains access to additional independent information and the statistic $Q(G(n,A))$ decreases. However, as the network grows larger, the expert at the center of the star becomes disproportionately influential. This concentration of connectivity amplifies indirect dependence across forecasts and increases the statistic $Q(G(n,A))$. As a result, aggregation efficiency is maximized at an interior network size, implying that there exists an optimal number of experts to consult when the communication structure among experts takes the form of a star.

The previous result characterizes a connected network structure that attains the largest asymptotic value of $Q(G(n,A))$ among sparse connected networks. It is therefore natural to compare it with networks at the opposite extreme in terms of degree heterogeneity. Because $d$-regular networks assign the same degree to every node (i.e., $d_i = d$ for all $i$), they provide a canonical benchmark for assessing how degree homogeneity affects the statistic $Q(G(n,A))$.

% ---------------------------------------------------------
\begin{proposition}
\label{prop:effic-d-reg}
Let $G(n,A)$ be a $d$-regular network, where $d_i=d\leq n$ for every $i$. Then
\[
Q(G(n,A)) = \frac{1}{n}.
\]
Consequently,
\[
\lim_{n \to \infty} Q(G(n,A)) = 0,
\]
which corresponds to the minimal asymptotic value attainable by $Q(G(n,A))$.
\end{proposition}
% ---------------------------------------------------------

In a $d$-regular network each node has the same degree. Increasing density raises the number of walks of length two between pairs of nodes, but it also increases the normalization term in exactly the same proportion. Because every node contributes symmetrically to the set of indirect connections captured by $A^2$, these two effects offset each other, implying that $Q(G(n,A)) = 1/n$ independently of the value of $d$ for every connected $d$-regular network..

From the perspective of information aggregation, Regular networks ensure that each expert’s signal enters the communication-induced data-generating process symmetrically. No expert becomes disproportionately central in the set of indirect connections across forecasts, and therefore no imbalance arises in how information propagates through the network. As a result, aggregation efficiency does not depend on network density within the class of connected regular networks. Instead, degree heterogeneity—rather than sparsity or density—is the key determinant of aggregation performance.

% ---------------------------------------------------------
\subsection{Local Perturbations of Extremal Networks}
\label{subsec:local-perturb}
% ---------------------------------------------------------

The previous results identify star and $d$-regular networks as extremal benchmark configurations for the statistic $Q(G(n,A))$, corresponding respectively to maximal degree heterogeneity within sparse connected networks and complete degree homogeneity. Most communication structures, however, lie between these two benchmarks. We therefore study how small deviations from these extremal topologies affect aggregation efficiency and how sensitive the statistic $Q(G(n,A))$ is to local changes in network structure.

In particular, we consider perturbations that preserve the size of the network and its connectivity, so that each expert continues to observe at least one other expert. These perturbations consist of local adjustments to the pattern of connections among experts, allowing us to isolate how marginal changes in degree heterogeneity affect the statistic $Q(G(n,A))$ while keeping the global structure of the network fixed.

% ---------------------------------------------------------
\begin{proposition}
\label{prop:rewire-cluster-star}
Let $G_s = G(n, A)$ be a star network on $n \geq 4$ nodes, where node $1$ is the center and nodes $2, 3, \ldots, n$ are peripheral nodes. Construct the following modified networks:
\begin{enumerate}
    \item $\Tilde{G}_s = G(n, \Tilde{A})$ by disconnecting one peripheral node from node $1$ and reconnecting it to another peripheral node (rewiring);
    \item $\Bar{G}_s = G(n, \Bar{A})$ by adding one edge between two peripheral nodes (clustering).
\end{enumerate}
Then, for all $n \geq 4$,
\[
Q(\Tilde{G}_s) - Q(G_s) < 0
\quad \text{and} \quad
Q(\Bar{G}_s) - Q(G_s) < 0.
\]
\end{proposition}
% ---------------------------------------------------------

Proposition~\ref{prop:rewire-cluster-star} shows that local perturbations that redistribute connectivity away from the center strictly decrease the statistic $Q(G(n,A))$. In this sense, the star network is locally maximal with respect to these edge modifications, reinforcing its role as the least efficient sparse connected topology for information aggregation.

We next consider analogous perturbations of $d$-regular networks. Since regular networks eliminate degree heterogeneity, it is natural to investigate whether small deviations from regularity increase the statistic $Q(G(n,A))$.

% ---------------------------------------------------------
\begin{proposition}[Local efficiency of $d$-regular networks]
\label{prop:delete-link-regular}
Let $G_d = G(n, A)$ be a $d$-regular network with $n \geq 4$ nodes. Consider two benchmark cases: (i) the complete network ($d=n$), and (ii) the cycle network ($d=3$). Let $\Tilde{G}_d = G(n, \Tilde{A})$ be the network obtained by deleting a single edge from $G_d$. Then, in both cases and for all $n \geq 4$,
\[
Q(\Tilde{G}_d) - Q(G_d) > 0.
\]
\end{proposition}
% ---------------------------------------------------------

That is, the statistic $Q(G(n,A))$ strictly increases under these local deletions of links. This shows that the minimal value of $Q(G(n,A))$ attained by $d$-regular networks is locally robust, in the sense that small departures from regularity increase the statistic $Q(G(n,A))$ and move the network away from its minimal asymptotic value.

Taken together, Propositions~\ref{prop:rewire-cluster-star} and~\ref{prop:delete-link-regular} show that the extremal properties of star and regular networks persist under local edge perturbations. In both cases, small deviations from these benchmark structures move the statistic $Q(G(n,A))$ toward the interior of its feasible range between the regular-network benchmark $1/n$ and the sparse network upper benchmark approaching $1/4$.

% -------------------------------------------------
\subsection{Multiple Connected Components}
\label{subsec:mult-conn-comp}
% -------------------------------------------------

The previous results characterize extremal aggregation efficiency within the class of connected networks. In many applications, however, experts may exchange information only within subgroups. We therefore extend the analysis to networks with multiple connected components. Because star and regular networks characterize the extremal benchmark values of $Q(G(n,A))$ within the class of connected sparse networks, we focus on unions of these structures.

% ---------------------------------------------------------
\begin{proposition}
\label{prop:union-regular-components-Q}

Let $G(n,A)$ be an undirected network with $K \ge 1$ connected components \, $C_1,\dots,C_K$, where component $k$ contains $m_k$ nodes and $\sum_{k=1}^K m_k = n$. If each component $C_k$ is internally $d_k$-regular, with possibly different degrees across components, then
\[
Q(G(n,A))=\frac{1}{n}.
\]
\end{proposition}
% ---------------------------------------------------------

This result extends the efficiency property of connected regular networks to the case of multiple components. Within each component, regularity implies that the number of walks of length two scales proportionally with the square of the common degree, so that numerator and denominator terms in the definition of $Q(G(n,A))$ cancel exactly. Therefore, the efficiency result for regular networks does not rely on connectivity but only on degree homogeneity.

We now consider the corresponding decomposition for star networks. Unlike the regular case, the statistic $Q(G(n,A))$ depends on the size distribution of the components. As a result, aggregation efficiency depends on how nodes are allocated across star components.

% ---------------------------------------------------------
\begin{proposition}
\label{prop:union-stars-Q}

Let $G(n,A)$ be an undirected network with $K\ge 1$ connected components. 
Assume each component $k$ is a star on $m_k\ge 3$ nodes, and $\sum_{k=1}^K m_k = n$.
Then
\[
Q(n,A)
=
\frac{1}{n^2}\sum_{k=1}^K T(m_k),
\qquad
T(m)
=
2-\frac{1}{m}+\frac{m(m-1)}{4}.
\]
\end{proposition}
% ---------------------------------------------------------

For large $m$, the leading term in $T(m)$ is $m^2/4$, so the expression admits the approximation
\[
Q(n,A)\approx \frac{1}{4n^2}\sum_{k=1}^K m_k^2
=
\frac{1}{4}\sum_{k=1}^K \left(\frac{m_k}{n}\right)^2.
\]
The right-hand side is proportional to the Herfindahl index of the component size shares. Holding $n$ fixed, it follows that $Q(n,A)$ increases when the mass of nodes is concentrated in fewer components, and decreases when the same mass is spread across more components.

A useful benchmark is the case of equally sized stars. If $m_k=n/K$ for all $k$, then the approximation becomes
\[
Q(n,A)\approx \frac{1}{4}\,K\left(\frac{1}{K}\right)^2=\frac{1}{4K},
\]
so increasing the number of star components decreases $Q(n,A)$ at rate $1/K$.

This complements the earlier result that the single connected star topology attains the largest asymptotic values of $Q(G(n,A))$ within the class of sparse networks on $n$ nodes. Indeed, when $K=1$ the present formula recovers the single-star case, whereas when $K>1$ the $n$ nodes are distributed across multiple star components with distinct centers, reducing the concentration of indirect information flows and therefore lowering $Q(n,A)$.

The previous results characterize aggregation efficiency across deterministic network structures and identify degree heterogeneity as the main driver of variation in the statistic $Q(G(n,A))$. We now examine how these insights extend to random network environments, where the topology itself is generated by a stochastic process.

% ========================================================
\section{Random Networks}
\label{sec:random-nets}
% ========================================================

We now study the behavior of the statistic $Q(G(n,A))$ in random network environments. Random graph models provide a natural benchmark for aggregation efficiency when the network structure is not fixed deterministically. They also allow us to characterize how the statistic $Q(G(n,A))$ behaves in typical large networks and to relate these outcomes to the extremal topologies analyzed in the previous section.

% ===========================================
\subsection{Poisson Random Graph Model} 
\label{subsec:poisson-rgm}
% ===========================================

We consider a Poisson random graph model in which $n$ experts may connect randomly with independent probability $p$. This model corresponds to the Erd\"os–R\'enyi random graph $G(n,p)$ augmented with deterministic self-loops. The adjacency matrix $A$ is generated as follows. For all $i$, $A_{ii}=1$. For $i\neq j$, the entries $A_{ij}=A_{ji}$ are independent Bernoulli random variables with parameter $p$. For any node $i$, the number of off-diagonal links follows a binomial distribution with mean $(n-1)p$, and since each node has a deterministic self-loop, the expected total degree equals $(n-1)p+1$.

Fixing $n$ and $p$ therefore induces a probability distribution over undirected networks with self-loops, under which every such network has strictly positive probability. We analyze how aggregation efficiency behaves asymptotically as the expected degree varies with network size.

% ---------------------------------------------------------
\begin{proposition}
\label{prop:poisson-3-regimes}
Under the Poisson random graph model $G(n,p)$, let $\lambda_n=(n-1)p_n$, with $p_n$ possibly depending on $n$. Then,
\begin{enumerate}
\item If $\lambda_n \to 0$ as $n\to\infty$, then
\[
\lim_{n\to\infty} n\,\mathbb E\!\left[Q(n,A)\right] = 1.
\]
\item If $\lambda_n \to \lambda \in (0,\infty)$ as $n\to\infty$, then there exists a constant $c(\lambda)\in(1,\infty)$ such that
\[
\lim_{n\to\infty} n\,\mathbb E\!\left[Q(n,A)\right] = c(\lambda).
\]
\item If $\lambda_n \to \infty$ as $n\to\infty$, then
\[
\lim_{n\to\infty} n\,\mathbb E\!\left[Q(n,A)\right] = 1.
\]
\end{enumerate}

\end{proposition}

Proposition~\ref{prop:poisson-3-regimes} implies that $\mathbb{E}[Q(n,A)] \sim c(\lambda)/n$. Thus,
\[
\left|\mathbb{E}[Q(n,A)]-\frac{1}{n}\right|
=
\frac{|c(\lambda)-1|}{n}
\to 0,
\]
so absolute deviations from the regular-network benchmark vanish asymptotically. However,
\[
n\left|\mathbb{E}[Q(n,A)]-\frac{1}{n}\right|
\to |c(\lambda)-1|,
\]
showing that degree heterogeneity generates a persistent relative distortion in aggregation efficiency.

The three regimes describe how aggregation efficiency depends on expected degree as the network grows. When $\lambda_n\to 0$, the network becomes sparse and most experts observe only their own forecasts, implying $n\,\mathbb{E}[Q(n,A)]\to 1$.

When $\lambda_n\to\infty$, degree dispersion becomes negligible relative to mean degree, and aggregation efficiency again approaches the regular-network benchmark.

The intermediate regime $\lambda_n\to\lambda\in(0,\infty)$ corresponds to networks with asymptotically constant expected degree. In this case, local degree fluctuations persist and generate the distortion factor $c(\lambda)>1$. Although $\mathbb{E}[Q(n,A)]\to 0$, convergence occurs more slowly than in dense networks because residual degree heterogeneity continues to affect aggregation efficiency.

\begin{figure}[H]
\centering

\begin{minipage}[c]{0.49\textwidth}
\centering
\includegraphics[width=\linewidth]{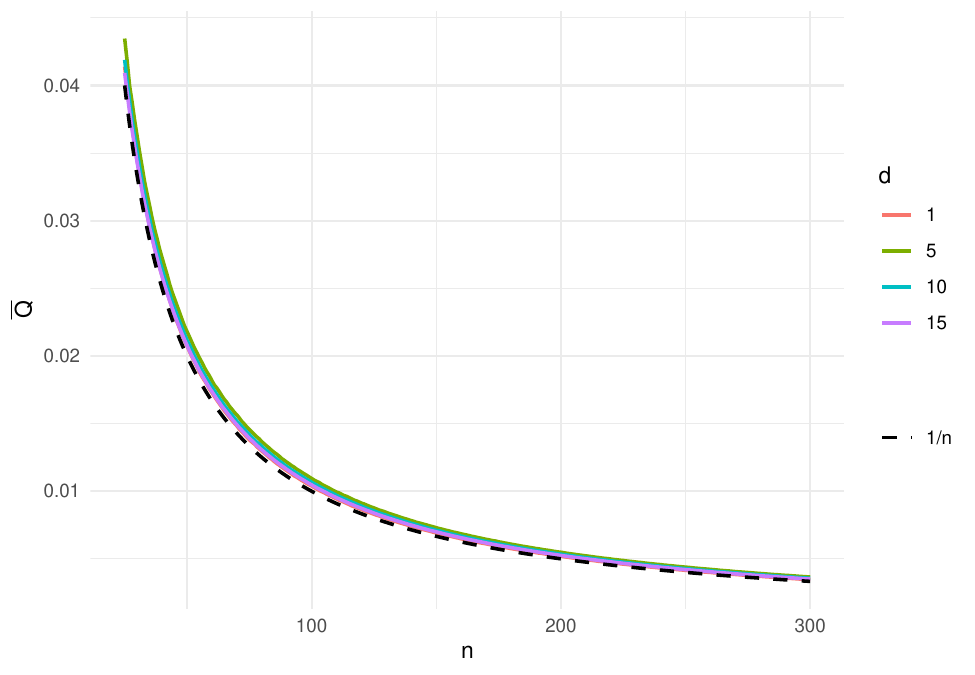}

\vspace{3pt}
{\footnotesize (a) Average $Q(G(n,A))$ ($\bar{Q}$) as a function of $n$, for different values of the average degree $d$}
\end{minipage}
\hfill
% --- Right: PDF plot ---
\begin{minipage}[c]{0.49\textwidth}
\centering
\includegraphics[width=\linewidth]{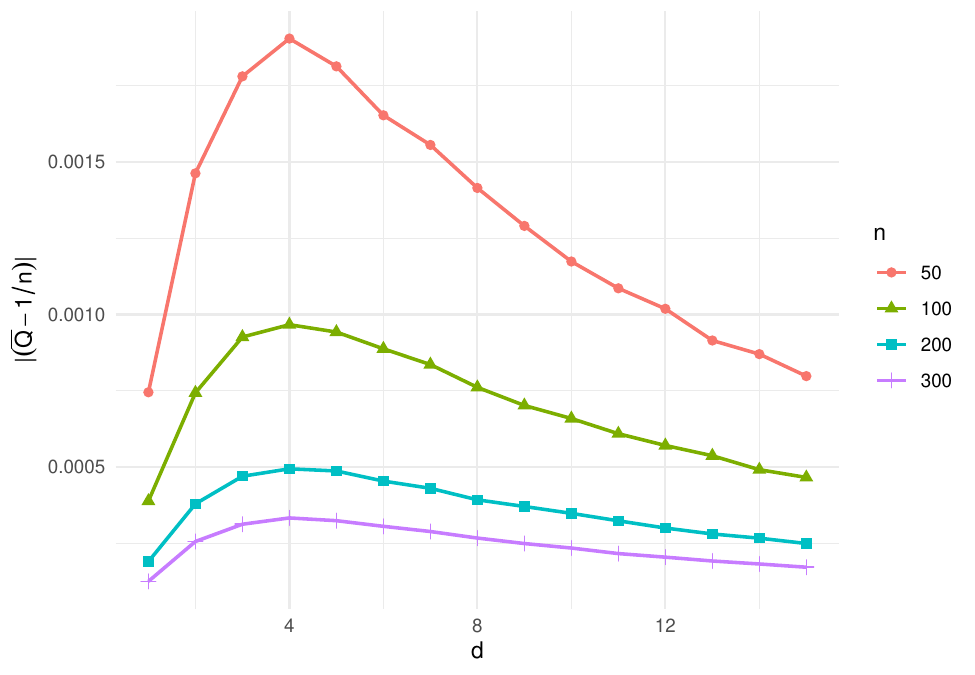}

\vspace{3pt}
{\footnotesize (b) $|\bar Q - 1/n|$ as a function of $d$, for different network sizes $n$}
\end{minipage}

\caption{Simulations of Poisson random graph model with $p=\frac{d}{n}$.}
\label{fig:sim-prgm}
\end{figure}

Figure~\ref{fig:sim-prgm} illustrates this intermediate regime using $p=d/n$, so that $\lambda_n\to d\in(0,\infty)$. Panel (a) shows that $\bar Q(G(n,A))$ decreases at rate $1/n$, consistent with Proposition~\ref{prop:poisson-3-regimes}. Panel (b) reports the deviation $|\bar Q(G(n,A)) - 1/n|$ as a function of $d$. These deviations shrink with $n$ but remain proportional to $(c(d)-1)/n$, confirming that residual degree heterogeneity affects aggregation efficiency when expected degree remains asymptotically constant.

We next consider a class of random networks in which degree heterogeneity persists asymptotically. These networks provide a natural benchmark for environments in which some experts accumulate disproportionately many informational connections. We study how this form of degree heterogeneity affects aggregation efficiency as the network grows.

% ==============================================
\subsection{Preferential Attachment Model}
\label{subsec:PA-model}
% ==============================================

In the Preferential Attachment model, the network grows through the successive addition of nodes. The initial network is assumed to be a complete graph with $m_0$ nodes. Each new node connects to $m \geq 1$ existing nodes, where the probability of selecting an existing node is proportional to its degree. If $N(t)$ denotes the set of nodes present at time $t$, the probability that a node with degree $d$ receives a link from a newly arriving node is
\[
p_t(d)=\frac{d}{\sum_{i\in N(t)} d_i}.
\]

For fixed $m$, the resulting network exhibits a scale-free degree distribution with power-law tail
\begin{equation}
\label{eq:deg-dist-PAmodel}
P(d)=\frac{2m(m+1)}{d(d+1)(d+2)}\sim d^{-3},
\quad \text{as } d\to\infty,
\end{equation}
which implies the persistent emergence of highly connected hubs as the network grows.

% ---------------------------------------------------------
\begin{proposition}
\label{prop:PA-Q-scaling}

Under the preferential attachment model with fixed parameter $m\ge 1$,

\[
\mathbb E[Q(n,A)]
=
\Theta\!\left(\frac{\log n}{n}\right)
\quad \text{as } n\to\infty.
\]

\end{proposition}
% ---------------------------------------------------------

Preferential attachment networks exhibit persistent degree heterogeneity and the emergence of hubs. These features increase the dispersion of the normalization terms appearing in $Q(G(n,A))$ and therefore reduce aggregation efficiency relative to regular networks. However, because preferential attachment does not generate the extreme concentration of influence observed in star networks, the statistic $Q(G(n,A))$ still converges to zero asymptotically. The convergence rate $\Theta(\log n/n)$ reflects the persistence of hub-driven heterogeneity as the network grows. Preferential attachment networks therefore occupy an intermediate position between regular and star networks in terms of aggregation efficiency, with $Q(G(n,A))$ decaying faster than in star networks but more slowly than in regular networks.\footnote{ Here $\Theta(\cdot)$ denotes asymptotic order in the usual sense. In particular, for two sequences $a_n$ and $b_n$, we write $a_n=\Theta(b_n)$ if there exist constants $c_1,c_2>0$ such that $c_1 b_n \le a_n \le c_2 b_n$ for all sufficiently large $n$.}

\begin{figure}[H]
\centering
\includegraphics[width=0.6\linewidth]{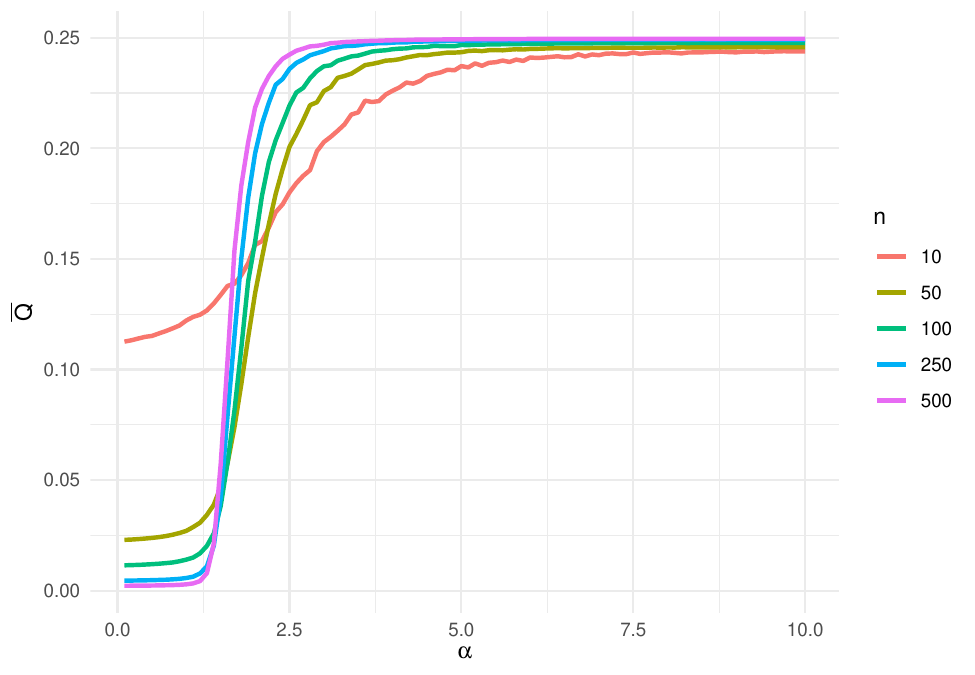}

\caption{Average $Q(G(n,A))$ ($\bar{Q}$) across simulations of the Preferential Attachment model with attachment probability $P(i)\propto d_i^{\alpha}+a$, with $a=1$}
\label{fig:pa-sim-alphas}
\end{figure}

Figure~\ref{fig:pa-sim-alphas} illustrates how aggregation efficiency varies when the strength of preferential attachment is modified. The simulations consider attachment probabilities of the form $P(i)\propto d_i^{\alpha}+a$ with $a=1$. When $\alpha=0$, links are formed uniformly at random and the network behaves similarly to the Poisson random graph benchmark. As $\alpha$ increases, attachment becomes increasingly concentrated on high-degree nodes, leading to stronger hub formation and larger values of $Q(G(n,A))$. For sufficiently large $\alpha$, the network approaches a star-like configuration and aggregation efficiency deteriorates accordingly. Consistent with Proposition~\ref{prop:PA-Q-scaling}, the statistic $Q(G(n,A))$ decreases with network size but at a slower rate than in the Poisson model due to the persistence of degree heterogeneity. Moreover, the simulated values remain within the theoretical asymptotic benchmark range $\frac{1}{n} \leq Q(G(n,A)) \leq \frac{1}{4}$, where the lower bound holds for all connected networks and the upper benchmark corresponds to sparse connected structures, with the star network attaining the largest asymptotic value within this class (Proposition~\ref{prop:bounds-Q-tree}). This confirms that increasing preferential attachment gradually shifts the network toward the regime of maximal aggregation distortion associated with hub-dominated sparse connected structures.

% ===========================================================
\subsection{Comparing Deterministic and Random Networks}
\label{subsec:comp-det-rand-net}
% ===========================================================

The results of this section complement the deterministic analysis of Section~\ref{sec:main-res}. Regular networks minimize the statistic $Q(G(n,A))$ asymptotically, while star networks attain the largest asymptotic values within the class of sparse networks. The Poisson random graph model approaches the regular network benchmark when expected degree grows with network size, whereas preferential attachment networks converge more slowly due to persistent degree heterogeneity.

Taken together, these results show that aggregation efficiency in large networks depends primarily on the extent of degree heterogeneity rather than on network density alone.

% =============================================
\section{Conclusion}
\label{sec:conclusion}
% =============================================

This paper studies how communication across experts prior to aggregation by the decision-maker affects the performance of forecast combination. Communication across experts is modeled as Bayesian updating along a fixed interaction network, so dependence across reported forecasts arises from information sharing. The analysis introduces the statistic $Q(G(n,A))$ as a measure of how network structure shapes the efficiency of information aggregation when reported forecasts become correlated through local interaction. The results show that aggregation efficiency depends primarily on the degree distribution of the network rather than on its density alone.

Within the class of connected networks, $d$-regular networks minimize $Q(G(n,A))$, while star networks attain the largest asymptotic values within the class of sparse networks. These extremal structures clarify how degree heterogeneity distorts the effective contribution of experts’ signals to the data-generating process observed by the decision-maker. In regular networks, signals are incorporated symmetrically through indirect information paths, whereas in star networks the central expert becomes disproportionately influential, generating the largest aggregation distortions among sparse connected structures. More generally, the statistic $Q(G(n,A))$ provides a tractable representation of how indirect information paths created by communication reshape the effective signal structure observed by the decision-maker, allowing aggregation efficiency to be compared across arbitrary network topologies.

Random network benchmarks complement these deterministic results. In the Poisson random graph model, aggregation efficiency converges to the regular network benchmark when expected number of off-diagonal links vanishes or diverges with network size, while persistent degree heterogeneity in the constant-degree regime generates a strictly larger asymptotic distortion factor relative to the regular network benchmark. Preferential attachment networks exhibit slower convergence due to persistent hub-driven degree heterogeneity, yielding intermediate behavior between regular networks and star networks within the sparse-network benchmark. Taken together, these results establish a hierarchy of aggregation efficiency across network classes indexed by their degree heterogeneity.

These findings contribute to the forecast combination literature by showing that interaction among experts prior to aggregation can substantially affect the informational content of pooled forecasts even when experts are unbiased.

Several extensions follow naturally from the present framework. First, allowing for repeated communication among experts would make it possible to study how consensus formation with heterogeneous signal precisions affects aggregation efficiency over time and whether dynamic interaction amplifies or mitigates distortions generated by degree heterogeneity. Second, introducing biased or strategically distorted signals would allow the interaction between network structure and misinformation to be characterized within the same framework. Third, allowing the decision-maker to condition explicitly on the network would make it possible to characterize optimal aggregation when reported forecasts are correlated through prior communication among experts. Finally, the results suggest treating communication structure itself as an object of analysis: one natural direction is to study the design of networks that minimize aggregation distortions subject to connectivity or communication constraints, while a complementary direction would study identification and robustness questions within environments where the communication network is not observed by the decision-maker.

Overall, the analysis highlights that the informational performance of forecast aggregation depends fundamentally on how experts communicate prior to reporting their forecasts, providing a natural bridge between the forecast combination literature and models of social learning in networks.

\pagebreak
%============================================================
% REFERENCES / BIBLIOGRAPHY
%============================================================
\clearpage
\nocite*
\bibliographystyle{ecta}
\bibliography{CCF_references_fv}

%============================================================
% APPENDIX
%============================================================
\clearpage
\appendix

% =======================================================
\section{Proofs of main results}
\label{append:proofs-main-res}
% =======================================================

% ==========================================================
\begin{proof}[\textbf{Proof of Proposition~\ref{prop:equiv-rules}}]

The posterior mean for the decision-maker is
\[
\Tilde{\theta}_0(\bm{x}) = \frac{\bm{1}' \bm{\Sigma}^{-1} \bm{x}}{\bm{1}' \bm{\Sigma}^{-1} \bm{1}}.
\]
According to Lemma ~\ref{lem:cov-mat-inv}, the numerator of the expression above becomes
\begin{align*}
\begin{split}
\bm{1}' \bm{\Sigma}^{-1} \bm{x} &= \frac{1}{\sigma^2(1 - \rho)} \bm{1}^{\prime}\left[ \bm{I} - \frac{\rho}{1 + (n - 1)\rho} \bm{1} \bm{1}' \right] \bm{x} \\
                                 &= \frac{1}{\sigma^2(1 - \rho)} \left[ \bm{1}^{\prime}\bm{I}\bm{x} - \frac{\rho}{1 + (n - 1)\rho} \bm{1}^{\prime}\bm{1} \bm{1}' \bm{x} \right]\\
                                 &= \frac{1}{\sigma^2(1 - \rho)} \left[ 1 - \frac{n\rho}{1 + (n - 1)\rho} \right] \bm{1}^{\prime}\bm{x}\\
                                 &= \frac{\bm{1}^{\prime}\bm{x}}{\sigma^2\left[1+(n-1)\rho\right]}  
\end{split}
\end{align*}

and the denominator
\begin{align*}
\begin{split}
\bm{1}' \bm{\Sigma}^{-1} \bm{1} &= \frac{1}{\sigma^2(1 - \rho)}\bm{1}^{\prime}\left[ \bm{I} - \frac{\rho}{1 + (n - 1)\rho} \bm{1} \bm{1}' \right] \bm{1} \\
                                &= \frac{1}{\sigma^2(1 - \rho)} \left[\bm{1}'\bm{1} - \frac{(\bm{1}'\bm{1})^2\rho}{1 + (n - 1)\rho}\right] \\
                                &= \frac{\bm{1}'\bm{1}}{\sigma^2\left[1+(n-1)\rho\right]}. 
\end{split}
\end{align*}
Thus,
\begin{equation}
\label{eq:post-mean-dm}
\Tilde{\theta}_0(\bm{x}) = \left(\bm{1}' \bm{1}\right)^{-1}\bm{1}' \bm{x} = \frac{1}{n}\sum_{i=1}^{n}x_i,    
\end{equation}
 and
\begin{equation}
\label{eq:post-var-dm}
\Tilde{\sigma}_{0}^{2} = \left(\bm{1}' \bm{\Sigma}^{-1} \bm{1}\right)^{-1} = \frac{\sigma^2}{n}[1+ (n-1)\rho].    
\end{equation}

Now consider any expert $i \in N$, who observes a subset of forecasts $\bm{x}_{N_i}$ with size $d_i$. The covariance matrix over $N_i$ is
\[
\bm{\Sigma}_{N_i} = \sigma^2 \left[ (1 - \rho)\bm{I}_{d_i} + \rho \bm{1}_{d_i} \bm{1}_{d_i}' \right],
\] and its inverse is
\[
\bm{\Sigma}_{N_i}^{-1} = \frac{1}{\sigma^2(1 - \rho)} \left[ \bm{I}_{N_i} - \frac{\rho}{1 + (d_i - 1)\rho} \bm{1}_{N_i} \bm{1}_{N_i}' \right].
\]
By direct calculation,
\[
\bm{1}_{N_i}' \bm{\Sigma}_{N_i}^{-1} \bm{x}_{N_i} = \frac{\bm{1}_{N_i}'\bm{x}_{N_i}}{\sigma^2[1+(d_i -1)\rho]},
\]
and the denominator is:
\[
\bm{1}_{N_i}' \bm{\Sigma}_{N_i}^{-1} \bm{1}_{N_i} = \frac{d_i}{\sigma^2(1 + (d_i - 1)\rho)}.
\]
Hence,
\begin{equation}
\label{eq:post-mean-exps}
\Tilde{\theta}_i(\bm{x}) = \left(\bm{1}_{N_i}'\bm{1}_{N_i}\right)^{-1}\bm{1}_{N_i}'\bm{x}_{N_i} =  \frac{1}{d_i} \sum_{j \in N_i} x_j,    
\end{equation} and
\begin{equation}
\label{eq:post-var-exps}
\Tilde{\sigma}^2_{i} = \left(\bm{1}_{N_i}' \bm{\Sigma}_{N_i}^{-1} \bm{1}_{N_i}\right)^{-1} = \frac{\sigma^2}{d_i}[1+(d_i - 1)\rho].    
\end{equation}

\end{proof}

% ==========================================================
\begin{proof}[\textbf{Proof of Proposition~\ref{prop:mean-var-ccf}}]

Equation~\eqref{eq:exp-mean-CCF} follows directly from linearity of expectations and the unbiasedness of experts’ forecasts, since the CCF estimator is a linear combination of the reported forecasts with weights summing to one. The variance of the CCF estimator is
\begin{align}
\begin{split}
\text{Var}\left(\Tilde{\theta}_0\left(\Tilde{\bm{\theta}}(\bm{x})\right) \mid G(n, A)\right) &= \text{Var}\left(\left(\bm{1}' \bm{1}\right)^{-1}\bm{1}' D^{-1}A\bm{x}\right) \\
                                 &= \frac{1}{n^2} \bm{1}'D^{-1}A\bm{\Sigma}A'D^{-1}\bm{1}\\
                                 &= \frac{1}{n^2} \bm{1}'D^{-1}A\left(\sigma^{2}\left[(1-\rho)\bm{I} + \rho \bm{1}\bm{1}'\right]\right)A'D^{-1}\bm{1} \quad \text{(by Lemma~\ref{lem:cov-mat-inv})} \\
                                 &= \frac{\sigma^2}{n^2} \left[(1-\rho)\bm{1}'D^{-1}A^2D^{-1}\bm{1} + \rho n^2\right] \quad \text{(by $A'=A$)} \\
                                 &= \sigma^2\left[\rho + (1-\rho) \frac{1}{n^2}\bm{1}'D^{-1}A^2D^{-1}\bm{1}\right] \\
                                 &= \sigma^2\left[\rho + (1-\rho) \frac{1}{n^2}\sum_{i=1}^{n}\sum_{j=1}^{n}\dfrac{\left(A^2\right)_{ij}}{d_i d_j}\right].
\end{split}
\end{align}

\end{proof}

% ======================================================
\begin{proof}[\textbf{Proof of Proposition~\ref{prop:bounds-Q-tree}}]
We first prove the lower bound. Define
\[
S_k=\sum_{i=1}^n \frac{A_{ik}}{d_i}.
\]
Using the definition of $Q(G(n,A))$ and the identity
\[
(A^2)_{ij}=\sum_{k=1}^n A_{ik}A_{kj},
\]
we can write
\[
Q(G(n,A))
=
\frac{1}{n^2}\sum_{k=1}^n S_k^2.
\]
Moreover,
\[
\frac{1}{n}\sum_{k=1}^n S_k
=
\frac{1}{n}\sum_{k=1}^n\sum_{i=1}^n\frac{A_{ik}}{d_i}
=
\frac{1}{n}\sum_{i=1}^n\frac{1}{d_i}\sum_{k=1}^n A_{ik}
=
\frac{1}{n}\sum_{i=1}^n 1
=
1.
\]
By Jensen's inequality,
\[
Q(G(n,A))
=
\frac{1}{n^2}\sum_{k=1}^n S_k^2
\geq
\frac{1}{n^2}\,n
\left(\frac{1}{n}\sum_{k=1}^n S_k\right)^2
=
\frac{1}{n}.
\]

We now derive the upper bound for tree networks using an edge-count argument. In any connected network with deterministic self-loops, the minimum degree is $d_{\min}=\min_i d_i=2$, corresponding to one self-loop and at least one neighbor, whereas the maximum degree is $d_{\max}=\max_i d_i=n$, corresponding to one self-loop and links to all other $n-1$ nodes. Hence,
\[
\frac{1}{n^2}\frac{1}{n^2}
\sum_{i=1}^n\sum_{j=1}^n (A^2)_{ij}
\leq
Q(G(n,A))
\leq
\frac{1}{4} \frac{1}{n^2}
\sum_{i=1}^n\sum_{j=1}^n (A^2)_{ij}.
\]
By Lemma~\ref{lem:sumpowerAsumd}, we re-write the inequality above as
\[
\frac{1}{n^4}\sum_{i=1}^n d_i^2
\leq
Q(G(n,A))
\leq
\frac{1}{4n^2}\sum_{i=1}^n d_i^2.
\]

Let $e$ denote the number of ordinary edges in the underlying simple graph obtained after removing self-loops. By Lemma~\ref{lem:boundsumsquaresdegrees-loops},
\[
\sum_{i=1}^n d_i^2
\leq
e\left(\frac{2e}{n-1}+n+2\right)+n.
\]
It follows that
\[
Q(G(n,A))
\leq
\frac{1}{4n^2}
\left[
e\left(\frac{2e}{n-1}+n+2\right)+n
\right].
\]

The right-hand side is increasing and convex in the edge count $e$. Thus, allowing denser connected graphs only weakens the implication of this inequality. A genuinely tight upper bound for arbitrary connected graphs would require exploiting more detailed topological information than the number of edges alone. We therefore focus on the minimally connected benchmark $e=n-1$, which delivers the strongest asymptotic consequence obtainable from this edge-count argument.

If the underlying simple graph is a tree, then $e=n-1$. Substituting this value into the previous inequality gives
\[
Q(G(n,A))
\leq
\frac{1}{4n^2}
\left[
(n-1)
\left(
\frac{2(n-1)}{n-1}+n+2
\right)
+n
\right].
\]
Simplifying further yields
\[
Q(G(n,A))
\leq
\frac{n^2+4n-4}{4n^2}
=
\frac{1}{4}+\frac{1}{n}-\frac{1}{n^2}.
\]

Combining the general lower bound with the tree-network upper bound yields
\[
\frac{1}{n}
\leq
Q(G(n,A))
\leq
\frac{1}{4}+\frac{1}{n}-\frac{1}{n^2}
\]
within the class of tree networks with deterministic self-loops. Taking limits gives
\[
0
\leq
\liminf_{n\to\infty} Q(G(n,A))
\leq
\limsup_{n\to\infty} Q(G(n,A))
\leq
\frac{1}{4},
\]
as claimed.
\end{proof}
% ======================================================

% ==========================================================
\begin{proof}[\textbf{Proof of Proposition~\ref{prop:ineff-star}}]
Assume node $1$ is the center of the star, and the peripheral (leaf) nodes are labeled $\{2, 3, \ldots, n\}$. The degree of the center is $d_1 = n$, accounting for its $n-1$ connections to the leaves and one self-loop. Each leaf node $i \in \{2, \ldots, n\}$ has degree $d_i = 2$, consisting of a connection to the center and a self-loop.

Let $A$ denote the adjacency matrix with self-loops included. Then the squared adjacency matrix $A^2$ has the following structure:
\begin{itemize}
    \item $(A^2)_{11} = n$, 
    \item $(A^2)_{ii} = 2$ for $i \geq 2$,
    \item $(A^2)_{1j} = (A^2)_{j1} = 2$ for $j \geq 2$,
    \item $(A^2)_{ij} = 1$ for $i \ne j \geq 2$ (leaf–leaf pairs).
\end{itemize}

We now compute $Q(G(n,A))$ by decomposing the sum into three components:

\medskip
\noindent\textbf{Diagonal terms:}
\[
\frac{(A^2)_{11}}{d_1^2} = \frac{n}{n^2} = \frac{1}{n}, \quad
\sum_{i=2}^n \frac{(A^2)_{ii}}{d_i^2} = (n - 1) \frac{2}{4} = \frac{n - 1}{2}.
\]

\noindent\textbf{Off-diagonal center–leaf terms:} there are $2(n - 1)$ off-diagonal terms of the form $2/2n$:
\[
\sum_{j=2}^n \left( \frac{(A^2)_{1j}}{d_1 d_j} + \frac{(A^2)_{j1}}{d_j d_1} \right) = \frac{2(n - 1)}{n}.
\]

\noindent\textbf{Off-diagonal leaf–leaf terms:} there are $(n - 1)(n - 2)$ such terms, each equal to $1/4$:
\[
\sum_{\substack{i, j = 2 \\ i \ne j}}^n \frac{(A^2)_{ij}}{d_i d_j} = \frac{(n - 1)(n - 2)}{4}.
\]

Summing all contributions, we obtain
\[
Q(n) = \frac{1}{n^2} \left( \frac{1}{n} + \frac{n - 1}{2} + \frac{2(n - 1)}{n} + \frac{(n - 1)(n - 2)}{4} \right).
\]

It follows directly from this expression that
\[
\lim_{n \to \infty} Q(G(n,A)) = \frac{1}{4}.
\]
\end{proof}

% ==========================================================

\begin{proof}[\textbf{Proof of Corollary~\ref{cor:star-15-min}}]
To characterize the minimizing value of $n$, we treat $Q(G(n,A))$ as a function defined on the positive real line and compute its derivative. This yields a unique interior critical point (when $n\geq3$) at
\[
n^* = 8 + 2\sqrt{13}.
\]
Since $n$ must be an integer, direct evaluation confirms that
\[
Q(15) < Q(14)
\quad \text{and} \quad
Q(15) < Q(16),
\]
so the minimizing network size is $n=15$.

% For a star network,
% \[
% \frac{dQ(G(n,A))}{dn} = \frac{n^2 - 16n + 12}{4n^4}.
% \]
% This expression is continuous and differentiable for $n > 0$, and the first-order condition yields the critical point
% \[
% n^{*} = 8 + 2\sqrt{13}.
% \]

% Since $Q(G(n,A))$ is strictly convex in $n$ for $n > 0$, this critical point corresponds to a unique global minimum. Finally, since $n$ must be an integer, the minimizer is given by
% \[
% n^{*} = \left\lfloor 8 + 2\sqrt{13} \right\rfloor = 15,
% \]
% where $\lfloor \cdot \rfloor$ denotes the floor function.

\end{proof}
% =================================================================

% =================================================================
\begin{proof}[\textbf{Proof of Proposition~\ref{prop:effic-d-reg}}]
Since the network is \(d\)-regular, every node has degree \(d\), so
\[
D=dI_n
\qquad\text{and}\qquad
D^{-1}=\frac{1}{d}I_n.
\]
Substituting into the expression for \(Q(G(n,A))\),
\[
Q(G(n,A))
=
\frac{1}{n^2}\,\bm{1}'\left(\frac{1}{d}I_n\right)A^2\left(\frac{1}{d}I_n\right)\bm{1}
=
\frac{1}{n^2d^2}\,\bm{1}'A^2\bm{1}.
\]

Now, because the graph is \(d\)-regular, each row of \(A\) sums to \(d\), i.e.,
\[
A\bm{1}=d\bm{1}.
\]
Multiplying by \(A\) once more,
\[
A^2\bm{1}
=
A(d\bm{1})
=
dA\bm{1}
=
d(d\bm{1})
=
d^2\bm{1}.
\]
Therefore,
\[
\bm{1}'A^2\bm{1}
=
\bm{1}'(d^2\bm{1})
=
d^2\,\bm{1}'\bm{1}
=
d^2 n.
\]
Substituting back,
\[
Q(G(n,A))
=
\frac{1}{n^2d^2}\cdot d^2 n
=
\frac{1}{n}.
\]

Hence \(Q(G(n,A))=1/n\) for any \(d\). Consequently,
\[
\lim_{n\to\infty}Q(G(n,A))
=
\lim_{n\to\infty}\frac{1}{n}
=
0.
\]
This corresponds to the minimal asymptotic value attainable by $Q(G(n,A))$ among connected networks.
\end{proof}
% =================================================================

% =================================================================
\begin{proof}[\textbf{Proof of Proposition~\ref{prop:rewire-cluster-star}}]

\textbf{PART 1: re-wiring a peripheral node.} 

For the original star network $G_s$, we have 
\[
d_1 = n, \quad d_i = 2 \quad \text{for all } i \geq 2.
\]
The entries of the squared adjacency matrix $A^2$ are
\begin{align*}
(A^2)_{11} &= n, \\
(A^2)_{ii} &= 2 \quad \text{for all } i \geq 2, \\
(A^2)_{1j} &= (A^2)_{j1} = 2 \quad \text{for all } j \geq 2, \\
(A^2)_{ij} &= 1 \quad \text{for all } i,j \geq 2, \text{ with } i\ne j.
\end{align*}

For the rewired network $\Tilde{G}_s$, we have
\[
\Tilde{d}_1 = n - 1, \quad \Tilde{d}_2 = 2, \quad \Tilde{d}_3 = 3, \quad \Tilde{d}_i = 2 \quad \text{for all } i \geq 4.
\]
The entries of the squared adjacency matrix $\Tilde{A}^2$ are:
\begin{align*}
(\Tilde{A}^2)_{11} &= n - 1, \\
(\Tilde{A}^2)_{ii} &= 
\begin{cases}
2 & \text{if } i = 2, \text{ or } i \geq 4, \\
3 & \text{if } i = 3,
\end{cases} \\
(\Tilde{A}^2)_{1j} &= (\Tilde{A}^2)_{j1} = 
\begin{cases}
1 & \text{if } j = 2, \\
2 & \text{if } j \geq 3,
\end{cases} \\
(\Tilde{A}^2)_{23} &= (\Tilde{A}^2)_{32} = 2, \\
(\Tilde{A}^2)_{2j} &= (\Tilde{A}^2)_{j2} = 0 \quad \text{for } j \geq 4, \\
(\Tilde{A}^2)_{3j} &= (\Tilde{A}^2)_{j3} = 1 \quad \text{for } j \geq 4, \\
(\Tilde{A}^2)_{ij} &= 1 \quad \text{for all } i,j \geq 4, \text{ with } i\ne j.
\end{align*}

We define the change in $Q$ as
\[
\Delta Q = Q(\Tilde{G}_s) - Q(G_s) = \frac{1}{n^2} \sum_{i=1}^{n} \sum_{j=1}^{n} \Delta_{ij},
\] where
\[
\Delta_{ij} = \left(\frac{(\Tilde{A}^2)_{ij}}{\Tilde{d}_i\Tilde{d}_j} - \frac{(A^2)_{ij}}{d_i d_j}\right).
\]

When $i=j$, we have
\[
\Delta_{ii} = \frac{1}{\Tilde{d}_i} - \frac{1}{d_i}  =
\begin{cases}
\frac{1}{n(n-1)} & \text{, if } i = 1 \\
0 & \text{, if } i = 2 \\
-\frac{1}{6} & \text{, if } i = 3 \\
0 & \text{, if } i \geq 4.
\end{cases} 
\]

When $i=1$, we have
\[
\Delta_{1j} =
\begin{cases}
\frac{1}{(n-1)} - \frac{1}{n} = \frac{1}{n(n-1)} & \text{, if } j = 1 \\
\frac{1}{2(n-1)} - \frac{2}{2n} = \frac{2-n}{2n(n-1)} & \text{, if } j = 2 \\
\frac{2}{3(n-1)} - \frac{2}{2n} = \frac{3-n}{3n(n-1)} & \text{, if } j = 3 \\
\frac{2}{2(n-1)} - \frac{2}{2n} = \frac{1}{n(n-1)} & \text{, if } j \geq 4 \\
\end{cases} 
\]

When $i=2$, we have
\[
\Delta_{2j} =
\begin{cases}
\Delta_{12} =  \frac{2-n}{2n(n-1)} & \text{, if } j = 1 \\
\frac{1}{2} - \frac{1}{2} = 0 & \text{, if } j = 2 \\
\frac{1}{3} - \frac{1}{4} = \frac{1}{12} & \text{, if } j = 3 \\
0 - \frac{1}{4} = - \frac{1}{4} & \text{, if } j \geq 4. \\
\end{cases} 
\]

When $i=3$, we have
\[
\Delta_{3j} =
\begin{cases}
\Delta_{13} = \frac{3-n}{3n(n-1)}  & \text{, if } j = 1 \\
\Delta_{23} = \frac{1}{12} & \text{, if } j = 2 \\
\frac{1}{3} - \frac{1}{2} = -\frac{1}{6}  & \text{, if } j = 3 \\
\frac{1}{6} - \frac{1}{4} = -\frac{1}{12}  & \text{, if } j \geq 4. \\
\end{cases} 
\]

Finally, when $i\geq 4$, we have
\[
\Delta_{ij} =
\begin{cases}
\Delta_{ji}  & \text{, if } j \leq 3 \\
0  & \text{, if } j \geq 4. \\
\end{cases} 
\]

To compute the total change, we split the sum into two parts: the contributions from $i,j \leq 3$, and those involving indices $i$ and $j \geq 4$.

For the first part,
\begin{align}
\label{eq:sum-3x3}
\sum_{i=1}^{3} \sum_{j=1}^{3} \Delta_{ij} &= \underbrace{\frac{1}{n(n-1)} + 0-\frac{1}{6}}_{\text{diagonal elements}} + \underbrace{2 \left(\frac{2-n}{2n(n-1)} + \frac{3-n}{3n(n-1)} + \frac{1}{12} \right)}_{\text{off-diagonal elements}} \nonumber \\
                    &= \frac{5}{3} \frac{3-n}{n(n-1)}.
\end{align}

Thus, for any $n\geq4$, $\sum_{i=1}^{3} \sum_{j=1}^{3}\Delta_{ij} <0$. For the second part, we have
\begin{align*}
\sum_{j\geq 4}^{n} \left(\Delta_{1j} + \Delta_{j,1} \right) &= 2 (n-3)\frac{1}{n(n-1)}, \\
\sum_{j\geq 4}^{n} \left(\Delta_{2j} + \Delta_{j,2} \right) &= - 2 (n-3) \frac{1}{4}, \\
\sum_{j\geq 4}^{n} \left(\Delta_{3j} + \Delta_{j,3} \right) &= - 2 (n-3) \frac{1}{12}, \\
\sum_{i\geq 4}^{n}\sum_{j\geq 4}^{n} \left(\Delta_{ij} + \Delta_{j,i} \right) &= 0.
\end{align*}

Summing all terms involving $j \geq 4$, we obtain
\begin{equation}
\label{eq:sum-off-3x3}
\sum_{i=1}^{n}\sum_{j=4}^{n}(\Delta_{ij}+\Delta_{ji}) = 2(n-3)\left[\frac{1}{n(n-1)} - \frac{1}{3}\right],    
\end{equation} which is negative for any $n\geq 4$.

Therefore, $\sum_{i=1}^{n}\sum_{j=1}^{n} \Delta_{ij} < 0$, for any $n\geq 4$, and therefore $\Delta Q = Q(\Tilde{G}_s) - Q(G_s) < 0$.

\vspace{2mm}
% PART II: CLUSTERING A STAR 
\paragraph{\textbf{PART 2: Clustering a star network}} Now we consider the case of clustering a star network by adding an edge between peripheral nodes 2 and 3.

In the clustered network $\Bar{G}_s$, we add an edge between leaves 2 and 3. Degrees become:
\[\bar{d}_1 = n, \quad \bar{d}_2 = \bar{d}_3 = 3, \quad \bar{d}_i = 2 \text{ for all } i \geq 4.\]
The squared adjacency matrix $\Bar{A}^2$ has entries:
\begin{align*}
(\Bar{A}^2)_{ii} &= 
\begin{cases}
n & \text{if } i = 1, \\
3 & \text{if } i \in \{2,3\}, \\
2 & \text{if } i \geq 4,
\end{cases} \\
(\Bar{A}^2)_{1j} &= (\Bar{A}^2)_{j1} = 
\begin{cases}
n & \text{if } j =1, \\
3 & \text{if } j \in \{2,3\}, \\
2 & \text{if } j \geq 4,
\end{cases} \\
(\Bar{A}^2)_{23} &= (\Bar{A}^2)_{32} = 3, \\
(\Bar{A}^2)_{2j} &= (\Bar{A}^2)_{j2} = 1 \text{ for } j \geq 4, \\
(\Bar{A}^2)_{3j} &= (\Bar{A}^2)_{j3} = 1 \text{ for } j \geq 4, \\
(\Bar{A}^2)_{ij} &= 1 \quad \text{for all } i,j \geq 4, \text{ with } i \ne j.
\end{align*}

Recap that
\[
\Delta Q = Q(\Bar{G}_s) - Q(G_s) = \frac{1}{n^2} \sum_{i=1}^{n} \sum_{j=1}^{n} \Delta_{ij}, \quad \text{where} \quad \Delta_{ij} = \left(\frac{(\Bar{A}^2)_{ij}}{\bar{d}_i \bar{d}_j} - \frac{(A^2)_{ij}}{d_i d_j}\right).
\]

For the diagonal terms ($i=j$), we get:
\begin{align*}
\Delta_{ii} = 
\begin{cases}
0 & i = 1, \\
\frac{1}{3} - \frac{1}{2} = -\frac{1}{6} & i=2,3, \\
0 & i \geq 4.
\end{cases}
\end{align*}

When $i=1$
\[
\Delta_{1j} = \Delta_{j1} = \frac{3}{3n} - \frac{2}{2n} = 0, \text{ for all } j.
\]

When $i=2$, we have
\[
\Delta_{2j} =
\begin{cases}
\Delta_{12} = 0 & \text{, if } j = 1 \\
\Delta_{22} = -\frac{1}{6} & \text{, if } j = 2 \\
\frac{1}{3} - \frac{1}{4} = \frac{1}{12} & \text{, if } j = 3 \\
\frac{1}{6} - \frac{1}{4} = - \frac{1}{12} & \text{, if } j \geq 4. \\
\end{cases} 
\]

When $i=3$, we have
\[
\Delta_{3j} =
\begin{cases}
\Delta_{13} = 0 & \text{, if } j = 1 \\
\Delta_{23} = \frac{1}{12} & \text{, if } j = 2 \\
\Delta_{33} = -\frac{1}{6}  & \text{, if } j = 3 \\
\frac{1}{6} - \frac{1}{4} = -\frac{1}{12}  & \text{, if } j \geq 4. \\
\end{cases} 
\]

Finally, when $i\geq 4$, we have
\[
\Delta_{ij} =
\begin{cases}
\Delta_{ji}  & \text{, if } j \leq 3 \\
0  & \text{, if } j \geq 4. \\
\end{cases} 
\]

Now we compute the full sum of changes
\begin{align*}
\sum_{i=1}^n \sum_{j=1}^n \Delta_{ij} &= \underbrace{2\left(\frac{1}{3} - \frac{1}{2}\right)}_{\text{entries (2,2) amd ((3,3)}} + \underbrace{2\left(\frac{1}{3} - \frac{1}{4}\right)}_{\text{entries (2,3) and (3,2)}} + \underbrace{4(n-3)\left(\frac{1}{6} - \frac{1}{4}\right)}_{\text{entries $(2,j)$, $(3,j)$, $(j,2)$ and $(j,3)$ for $j\geq4$}} \\
&= -\frac{1}{3} + \frac{1}{6} - \frac{1}{3}(n - 3) = -\frac{1 + 2(n - 3)}{6} = \frac{5- 2n}{6}.
\end{align*}

Therefore, $\Delta Q < 0$, and the clustering deviation strictly reduces $Q(G_s)$.

\end{proof}
% =================================================================

% =================================================================
\begin{proof}[\textbf{Proof of Proposition~\ref{prop:delete-link-regular}}]
The proof is divided into two parts, corresponding to each of the two network structures considered.

\paragraph{\textbf{Part 1: Complete Network.}}Let $G = G(n, A)$ be a complete network with $n \geq 4$ nodes. The adjacency matrix $A$ satisfies
\[
A_{ij} = 1 \quad \text{for all } i,j \in \{1, \dots, n\}.
\] and each node has degree
\[
d_i = \sum_{j=1}^n A_{ij} = n \quad \text{for all } i.
\]
Thus, the squared adjacency matrix is
\[
(A^2)_{ij} = \sum_{k=1}^n A_{ik} A_{kj} = n \quad \text{for all } i,j.
\]
Hence, each normalized entry is:
\[
\frac{(A^2)_{ij}}{d_i d_j} = \frac{n}{n^2} = \frac{1}{n} \quad \text{for all } i,j.
\]

Now let $\Tilde{G} = G(n, \Tilde{A})$ be the network obtained by deleting (without loss of generality with respect to which edge is deleted) the edge $(2,3)$ from $G$. The perturbed adjacency matrix $\Tilde{A}$ satisfies:
\[
\Tilde{A}_{ij} =
\begin{cases}
0 & \text{if } (i,j) \in \{(2,3), (3,2)\}, \\
1 & \text{otherwise}.
\end{cases}
\]
Node degrees become:
\[
\Tilde{d}_i =
\begin{cases}
n - 1 & \text{if } i \in \{2,3\}, \\
n & \text{otherwise}.
\end{cases}
\]

The entries of the squared adjacency matrix $\Tilde{A}^2$ satisfy:
\[
(\Tilde{A}^2)_{ij} =
\begin{cases}
n - 1 & \text{if } i = j \in \{2,3\}, \\
n - 2 & \text{if } (i,j) \in \{(2,3), (3,2)\}, \\
n & \text{if } i,j \notin \{2,3\}, \\
n - 1 & \text{if } i \in \{2,3\}, j \notin \{2,3\} \text{ or vice versa}.
\end{cases}
\]

Corresponding normalized values are:
\[
\frac{(\Tilde{A}^2)_{ij}}{\Tilde{d}_i \Tilde{d}_j} =
\begin{cases}
\frac{1}{n - 1} & \text{if } i = j \in \{2,3\}, \\
\frac{n - 2}{(n - 1)^2} & \text{if } (i,j) \in \{(2,3), (3,2)\}, \\
\frac{1}{n} & \text{otherwise}.
\end{cases}
\]    

Again, we define the change in $Q$ as
\[
\Delta Q = Q(\Tilde{G}) - Q(G) = \frac{1}{n^2} \sum_{i=1}^{n} \sum_{j=1}^{n} \Delta_{ij},
\] where
\[
\Delta_{ij} = \left(\frac{(\Tilde{A}^2)_{ij}}{\Tilde{d}_i\Tilde{d}_j} - \frac{(A^2)_{ij}}{d_i d_j}\right).
\]

For this case
\begin{align*}
\Delta Q &= \frac{1}{n^2}\left(2\left(\frac{1}{(n-1)}-\frac{1}{n}\right)+2\left(\frac{n-2}{(n-1)^2}-\frac{1}{n}\right)\right)  \\
            &= \frac{1}{n^2}\left(\frac{2(n-1)}{n(n-1)^2}-\frac{2}{n(n-1)^2}\right) \\
            &= \frac{2n-4}{n^3(n-1)^2}.
\end{align*}

Thus, $\Delta Q > 0$ for every $n\geq3$, so the perturbation strictly increases $Q(G(n,A))$.

\paragraph{\textbf{Part 2: cycle network.}} 
Let $G_c = G(n, A)$ be a cycle network on $n \geq 4$ nodes. Consider the network $G_\ell = G(n, A_\ell)$ obtained by deleting a single edge from $G_c$, thereby converting the cycle into a line network. For any line network, the adjacency matrix $A_\ell$ satisfies:
\[
(A_\ell)_{ij} =
\begin{cases}
1 & \text{if } i = j, \\
1 & \text{if } |i - j| = 1, \\
0 & \text{otherwise}.
\end{cases}
\]

The degree of each node is:
\[
d_i =
\begin{cases}
2 & \text{if } i \in \{1, n\}, \\
3 & \text{if } 2 \leq i \leq n - 1.
\end{cases}
\]

The entries of the squared adjacency matrix $A_\ell^2$ are:
\[
(A_\ell^2)_{ij} =
\begin{cases}
2 & \text{if } i \in \{1, n\} \text{ and } i = j, \\
3 & \text{if } 2 \leq i \leq n - 1 \text{ and } i = j, \\
2 & \text{if } |i - j| = 1, \\
1 & \text{if } |i - j| = 2, \\
0 & \text{otherwise}.
\end{cases}
\]

Hence, the normalized entries of $A_\ell^2$ are:
\[
\frac{(A^2)_{ij}}{d_i d_j} =
\begin{cases}
\frac{1}{2} & \text{if } i = j \in \{1,n\}, \\
\frac{1}{3} & \text{if } i = j \in \{2,\dots,n-1\}, \\
\frac{1}{3} & \text{if } |i-j| = 1 \text{ and } \{i,j\} = \{1,2\} \text{ or } \{n-1,n\}, \\
\frac{2}{9} & \text{if } |i-j| = 1 \text{ and } i,j \in \{2,\dots,n-1\}, \\
\frac{1}{6} & \text{if } |i-j| = 2 \text{ and } \{i,j\} = \{1,3\} \text{ or } \{n-2,n\}, \\
\frac{1}{9} & \text{if } |i-j| = 2 \text{ and } i,j \in \{2,\dots,n-1\}, \\
0 & \text{otherwise}.
\end{cases}
\]

In this case,
\begin{align*}
Q_{\ell} & = \frac{1}{n^2}\left( \left(2 \frac{1}{2} + (n-2)\frac{1}{3}\right) + 2\left(2\frac{1}{3} + (n-3)\frac{2}{9}\right) +2\left(2\frac{1}{6}+(n-4)\frac{1}{9}\right)\right) \\
    &= \frac{1}{n^2}\left(\frac{1}{9} + n \right) \\
    &= \frac{1}{9n^2} + \frac{1}{n}.
\end{align*}

From Proposition~\ref{prop:effic-d-reg}, we have that $Q_c = \frac{1}{n}$. Therefore,
\[
\Delta Q = Q_{\ell} - Q_c = \frac{1}{9n^2} + \frac{1}{n} - \frac{1}{n} = \frac{1}{9n^2},
\] which implies that $\Delta Q > 0$ for all $n$.

\end{proof}
% =================================================================

% =================================================================
\begin{proof}[\textbf{Proof of Proposition~\ref{prop:union-regular-components-Q}}]
Since the network is the disjoint union of the components $C_1,\dots,C_K$, after relabeling nodes if necessary we may write $A$ in block-diagonal form as
\[
A=\mathrm{diag}(A_1,\dots,A_K),
\]
where $A_k$ is the $m_k\times m_k$ adjacency matrix of the subgraph induced by $C_k$. Similarly, $D$ is block diagonal,
\[
D=\mathrm{diag}(d_1 I_{m_1},\, d_2 I_{m_2},\,\dots,\, d_K I_{m_K}),
\]
where $I_{m_k}$ is the $m_k\times m_k$ identity matrix and, by assumption, all nodes in component $k$ have degree $d_k$.

Because $A$ and $D$ are block diagonal, so is $D^{-1}A^2D^{-1}$, and therefore
\[
\bm{1}'D^{-1}A^2D^{-1}\bm{1}
=
\sum_{k=1}^K
\bm{1}_{m_k}'\left(\frac{1}{d_k^2}A_k^2\right)\bm{1}_{m_k},
\]
where $\bm{1}_{m_k}$ is the $m_k\times 1$ all-ones vector. Since component $k$ is $d_k$-regular, each row of $A_k$ sums to $d_k$, implying
\[
A_k\bm{1}_{m_k}=d_k\bm{1}_{m_k}
\quad\Rightarrow\quad
A_k^2\bm{1}_{m_k}=d_k^2\bm{1}_{m_k}.
\]
Hence,
\[
\bm{1}_{m_k}'A_k^2\bm{1}_{m_k}
=
\bm{1}_{m_k}'(d_k^2\bm{1}_{m_k})
=
d_k^2\,\bm{1}_{m_k}'\bm{1}_{m_k}
=
d_k^2 m_k,
\]
and the contribution of component $k$ becomes
\[
\bm{1}_{m_k}'\left(\frac{1}{d_k^2}A_k^2\right)\bm{1}_{m_k}
=
\frac{1}{d_k^2}\, d_k^2 m_k
=
m_k.
\]
Summing across components yields
\[
\bm{1}'D^{-1}A^2D^{-1}\bm{1}
=
\sum_{k=1}^K m_k
=
n,
\]
and therefore
\[
Q(G(n,A))=\frac{1}{n^2}\,\bm{1}'D^{-1}A^2D^{-1}\bm{1}
=\frac{1}{n^2}\cdot n
=\frac{1}{n}.
\]
\end{proof}
% =================================================================

% =================================================================
\begin{proof}[\textbf{Proof of Proposition~\ref{prop:union-stars-Q}}]
Since the network is the disjoint union of its $K$ components, after relabeling nodes if necessary we may write the adjacency matrix in block-diagonal form as
\[
A=\mathrm{diag}\!\big(A_1,\dots,A_K\big),
\qquad
A^2=\mathrm{diag}\!\big(A_1^2,\dots,A_K^2\big).
\]
Hence, in the definition
\[
Q(n,A)=\frac{1}{n^2}\sum_{i=1}^n\sum_{j=1}^n \frac{(A^2)_{ij}}{d_i d_j},
\]
all cross-component terms vanish and we can decompose
\[
Q(n,A)
=
\frac{1}{n^2}\sum_{k=1}^K 
\sum_{i,j\in C_k}
\frac{(A_k^2)_{ij}}{d_i d_j}.
\]
We now compute the within-component contribution for a generic star component of size $m\ge 2$.

Fix a component of size $m$ and order its nodes so that node $1$ is the center and nodes $2,\dots,m$ are leaves.
Let $\bm{1}_{m-1}$ be the $(m-1)\times 1$ all-ones vector and $I_{m-1}$ the $(m-1)\times(m-1)$ identity matrix.
With self-loops and star edges, the adjacency matrix of this component is
\[
A_\star(m)
=
\begin{pmatrix}
1 & \bm{1}_{m-1}' \\
\bm{1}_{m-1} & I_{m-1}
\end{pmatrix}.
\]
Multiplying blocks gives
\[
A_\star(m)^2
=
\begin{pmatrix}
1+\bm{1}_{m-1}'\bm{1}_{m-1} & \bm{1}_{m-1}'+\bm{1}_{m-1}'I_{m-1} \\
\bm{1}_{m-1}+I_{m-1}\bm{1}_{m-1} & \bm{1}_{m-1}\bm{1}_{m-1}'+I_{m-1}^2
\end{pmatrix}
=
\begin{pmatrix}
m & 2\bm{1}_{m-1}' \\
2\bm{1}_{m-1} & \bm{1}_{m-1}\bm{1}_{m-1}'+I_{m-1}
\end{pmatrix}.
\]
Thus, in this star component, $(A_\star(m)^2)_{11}=m$, $(A_\star(m)^2)_{1\ell}=(A_\star(m)^2)_{\ell 1}=2$ for every leaf $\ell\in\{2,\dots,m\}$, and for leaves $\ell,\ell'\in\{2,\dots,m\}$ one has
\[
(A_\star(m)^2)_{\ell\ell}=2,
\qquad
(A_\star(m)^2)_{\ell\ell'}=1 \ \ \text{if }\ell\neq \ell'.
\]
Degrees in this component are
\[
d_1=m
\quad\text{(center: self-loop plus $m-1$ leaves),}
\qquad
d_\ell=2 \ \ \text{for each leaf }\ell\ge 2
\quad\text{(self-loop plus the center).}
\]

Define the within-component sum
\[
T(m)
:=
\sum_{i,j=1}^m \frac{(A_\star(m)^2)_{ij}}{d_i d_j}.
\]
We compute $T(m)$ by splitting ordered pairs $(i,j)$ into cases.

\medskip
\noindent\textit{(a) Center--center.}
There is one ordered pair $(1,1)$:
\[
\frac{(A_\star(m)^2)_{11}}{d_1 d_1}
=
\frac{m}{m^2}
=
\frac{1}{m}.
\]

\medskip
\noindent\textit{(b) Center--leaf and leaf--center.}
There are $m-1$ ordered pairs $(1,\ell)$ and $m-1$ ordered pairs $(\ell,1)$, each with numerator $2$ and denominator $m\cdot 2$:
\[
\frac{2}{m\cdot 2}=\frac{1}{m}.
\]
Hence the total contribution is
\[
(m-1)\cdot\frac{1}{m}+(m-1)\cdot\frac{1}{m}
=
\frac{2(m-1)}{m}.
\]

\medskip
\noindent\textit{(c) Leaf--leaf.}
There are $(m-1)$ diagonal ordered pairs $(\ell,\ell)$ with numerator $2$ and denominator $2\cdot 2=4$, giving contribution
\[
(m-1)\cdot\frac{2}{4}=\frac{m-1}{2}.
\]
There are $(m-1)(m-2)$ off-diagonal ordered pairs $(\ell,\ell')$ with $\ell\neq \ell'$, each with numerator $1$ and denominator $4$, giving contribution
\[
(m-1)(m-2)\cdot\frac{1}{4}=\frac{(m-1)(m-2)}{4}.
\]

\medskip
Summing (a)--(c) yields
\begin{align*}
T(m)
&=
\frac{1}{m}
+\frac{2(m-1)}{m}
+\frac{m-1}{2}
+\frac{(m-1)(m-2)}{4} \\
&=
\frac{2m-1}{m}
+
\frac{2(m-1)+(m-1)(m-2)}{4} \\
&=
\left(2-\frac{1}{m}\right)
+
\frac{(m-1)m}{4}
=
2-\frac{1}{m}+\frac{m(m-1)}{4}.
\end{align*}

Returning to the full graph with $K$ star components of sizes $m_1,\dots,m_K$, block-diagonality implies
\[
\sum_{i=1}^n\sum_{j=1}^n \frac{(A^2)_{ij}}{d_i d_j}
=
\sum_{k=1}^K T(m_k),
\]
and therefore
\[
Q(n,A)=\frac{1}{n^2}\sum_{k=1}^K T(m_k)
=
\frac{1}{n^2}\sum_{k=1}^K\left(2-\frac{1}{m_k}+\frac{m_k(m_k-1)}{4}\right).
\]

\end{proof}
% =================================================================

The proof of Proposition~\ref{prop:poisson-3-regimes} below relies on bounded second moments of the degree distribution under the self-loop normalization, which ensures uniform integrability of the sequence considered below.

% =================================================================
\begin{proof}[\textbf{Proof of Proposition~\ref{prop:poisson-3-regimes}}]
Define, again, for each $k\in\{1,\dots,n\}$,
\[
S_k \;=\; \sum_{i=1}^n \frac{A_{ik}}{d_i}.
\]
We first rewrite the double sum in $Q(n,A)$ in terms of $\{S_k\}_{k=1}^n$.
Since
\[
(A^2)_{ij}=\sum_{k=1}^n A_{ik}A_{kj},
\]
we have
\begin{align*}
\sum_{i=1}^n\sum_{j=1}^n \frac{(A^2)_{ij}}{d_i d_j}
&=
\sum_{i=1}^n\sum_{j=1}^n \frac{1}{d_i d_j}\sum_{k=1}^n A_{ik}A_{kj} \\
&=
\sum_{k=1}^n \sum_{i=1}^n\sum_{j=1}^n \frac{A_{ik}A_{kj}}{d_i d_j} \\
&=
\sum_{k=1}^n
\left(\sum_{i=1}^n \frac{A_{ik}}{d_i}\right)
\left(\sum_{j=1}^n \frac{A_{kj}}{d_j}\right).
\end{align*}
Because $A$ is symmetric, $A_{kj}=A_{jk}$, and therefore
\[
\sum_{j=1}^n \frac{A_{kj}}{d_j}
=
\sum_{j=1}^n \frac{A_{jk}}{d_j}
=
\sum_{i=1}^n \frac{A_{ik}}{d_i}
=
S_k.
\]
Thus the previous display becomes
\[
\sum_{i=1}^n\sum_{j=1}^n \frac{(A^2)_{ij}}{d_i d_j}
=
\sum_{k=1}^n S_k^2,
\]
and hence
\[
Q(n,A)=\frac{1}{n^2}\sum_{k=1}^n S_k^2.
\]

\medskip
\noindent\textbf{Step 1: exchangeability.}
The Poisson random graph model is invariant under relabeling of nodes: if $\pi$ is any permutation of $\{1,\dots,n\}$ and we relabel nodes by $\pi$, the resulting random adjacency matrix has the same distribution.
Since $S_k$ is a measurable function of $(A_{ij})_{i,j}$ and depends on $k$ only through the label of the column/row being referenced, this invariance implies that the random variables $S_1,\dots,S_n$ are identically distributed. In particular,
\[
\mathbb E[S_k^2] = \mathbb E[S_1^2]
\qquad\text{for all }k=1,\dots,n.
\]
Taking expectations in $Q(n,A)=\frac{1}{n^2}\sum_{k=1}^n S_k^2$ yields
\begin{align*}
n\,\mathbb E[Q(n,A)]
&=
n\,\mathbb E\!\left[\frac{1}{n^2}\sum_{k=1}^n S_k^2\right]
=
\frac{1}{n}\sum_{k=1}^n \mathbb E[S_k^2] \\
&=
\frac{1}{n}\sum_{k=1}^n \mathbb E[S_1^2]
=
\mathbb E[S_1^2].
\end{align*}

\medskip
\noindent\textbf{Step 2: a deterministic averaging identity.}
We claim that
\[
\frac{1}{n}\sum_{k=1}^n S_k = 1
\]
holds deterministically for every realized adjacency matrix $A$.
Indeed,
\begin{align*}
\sum_{k=1}^n S_k
&=
\sum_{k=1}^n \sum_{i=1}^n \frac{A_{ik}}{d_i}
=
\sum_{i=1}^n \frac{1}{d_i}\sum_{k=1}^n A_{ik}
=
\sum_{i=1}^n \frac{d_i}{d_i}
=
n,
\end{align*}
where we used $\sum_{k=1}^n A_{ik}=d_i$ by definition of degree.
Dividing by $n$ gives the claim, and taking expectations implies
\[
\mathbb E[S_1]=\frac{1}{n}\sum_{k=1}^n \mathbb E[S_k]=1.
\]
Therefore,
\[
\mathbb E[S_1^2]
=
\mathrm{Var}(S_1)+\big(\mathbb E[S_1]\big)^2
=
\mathrm{Var}(S_1)+1.
\]
Combining with Step 1 yields the identity
\[
n\,\mathbb E[Q(n,A)] = 1 + \mathrm{Var}(S_1).
\]

\medskip
We now analyze $\mathrm{Var}(S_1)$ in the three cases.

%------------------------------------------------------------
\medskip
\noindent\textbf{Case (1) $\lambda_n\to 0$. (sparse network)}
Let $N_1=\sum_{j\neq 1} A_{1j}$ be the off-diagonal degree of node $1$. Then $N_1\sim\mathrm{Bin}(n-1,p_n)$ and
\[
\Pr(N_1=0)=(1-p_n)^{n-1}\to 1
\qquad(\text{since }(n-1)p_n=\lambda_n\to 0).
\]
On the event $\{N_1=0\}$, node $1$ has no off-diagonal neighbors, so $d_1=1$ and $A_{i1}=0$ for all $i\neq 1$, while $A_{11}=1$. Hence
\[
S_1=\sum_{i=1}^n \frac{A_{i1}}{d_i}=\frac{A_{11}}{d_1}=1
\qquad\text{on }\{N_1=0\}.
\]
Therefore $S_1\to 1$ in probability, which implies $\mathrm{Var}(S_1)\to 0$, and thus
\[
n\,\mathbb E[Q(n,A)] = 1+\mathrm{Var}(S_1)\to 1.
\]

%------------------------------------------------------------
\medskip
\noindent\textbf{(iii) $\lambda_n\to\infty$.}
Let $\mu_n=1+\lambda_n=1+(n-1)p_n$.
For any fixed node $i$, the off-diagonal degree $\sum_{j\neq i}A_{ij}\sim\mathrm{Bin}(n-1,p_n)$, hence
\[
d_i=1+\mathrm{Bin}(n-1,p_n),
\qquad
\frac{d_i}{\mu_n}\to 1
\quad\text{in probability}.
\]
Write
\[
S_1
=
\sum_{i=1}^n \frac{A_{i1}}{d_i}
=
\frac{1}{\mu_n}\sum_{i=1}^n A_{i1}
+
\sum_{i=1}^n A_{i1}\left(\frac{1}{d_i}-\frac{1}{\mu_n}\right).
\]
The first term equals $d_1/\mu_n$ and converges to $1$ in probability because $d_1/\mu_n\to 1$.
For the second term, note that $\sum_{i=1}^n A_{i1}=d_1=O_p(\mu_n)$ while, for a typical neighbor $i$ of node $1$, $1/d_i-1/\mu_n=o_p(1/\mu_n)$ as degrees concentrate around $\mu_n$. Hence the remainder term is $o_p(1)$, so $S_1\to 1$ in probability. Therefore $\mathrm{Var}(S_1)\to 0$ and
\[
n\,\mathbb E[Q(n,A)] = 1+\mathrm{Var}(S_1)\to 1.
\]

%------------------------------------------------------------
\medskip
\noindent\textbf{(ii) $\lambda_n\to\lambda\in(0,\infty)$.}
In the constant-degree regime, degrees do not concentrate, and $S_1$ remains non-degenerate.
We formalize this using the standard local weak limit for sparse Poisson random graphs: the neighborhood of a uniformly random node converges in distribution to a Galton--Watson tree with offspring distribution $\mathrm{Pois}(\lambda)$.

Let $N\sim\mathrm{Pois}(\lambda)$ denote the limiting off-diagonal degree of node $1$, so that the total degree of node $1$ is $1+N$.
Moreover, conditional on being a neighbor of node $1$, a neighbor's total degree has the size-biased limit
\[
D^\star \;=\; 2 + \mathrm{Pois}(\lambda),
\]
corresponding to one self-loop, the edge to node $1$, and an asymptotically independent number of additional off-diagonal neighbors.
Let $X_1,X_2,\dots$ be i.i.d.\ copies of $1/D^\star$, independent of $N$, and define
\[
S^{(\lambda)} \;=\; \frac{1}{1+N} \;+\; \sum_{r=1}^{N} X_r.
\]
Under the local weak limit, $S_1 \Rightarrow S^{(\lambda)}$ as $n\to\infty$.
In addition, in this regime the degrees have bounded second moments, which implies $\{S_1^2\}_n$ is uniformly integrable; therefore convergence in distribution yields convergence of second moments:
\[
\mathbb E[S_1^2] \;\to\; \mathbb E\!\left[\left(S^{(\lambda)}\right)^2\right] \;=:\; c(\lambda).
\]
By Step 1, $n\,\mathbb E[Q(n,A)]=\mathbb E[S_1^2]$, hence $n\,\mathbb E[Q(n,A)]\to c(\lambda)$.

Finally, $c(\lambda)>1$. Indeed, $S^{(\lambda)}$ is non-degenerate: for example,
\[
\Pr(N=0)=e^{-\lambda}>0 \ \Rightarrow\ S^{(\lambda)}=1,
\]
while also
\[
\Pr(N=1)\Pr(\mathrm{Pois}(\lambda)=1)>0 \ \Rightarrow\ S^{(\lambda)}=\frac12+\frac13\neq 1,
\]
so $\Pr(S^{(\lambda)}\neq 1)>0$ and hence $\mathrm{Var}(S^{(\lambda)})>0$.
Therefore,
\[
c(\lambda)=\mathbb E\!\left[\left(S^{(\lambda)}\right)^2\right]
=
1+\mathrm{Var}(S^{(\lambda)})
\in (1,\infty).
\]

\medskip
Combining (i)--(iii) completes the proof.
\end{proof}
% =================================================================

% -------------------------------------------------------------
\begin{proof}[\textbf{Proof of Proposition~\ref{prop:PA-Q-scaling}}]

Let
\[
S_k=\sum_{i=1}^n\frac{A_{ik}}{d_i}.
\]
and
\[
Q(n,A)=\frac{1}{n^2}\sum_{k=1}^n S_k^2.
\]
Let $k$ denote a node drawn uniformly at random from $\{1,\ldots,n\}$. Then
\[
\mathbb E[Q(n,A)]
=
\frac{1}{n}\mathbb E[S_k^2].
\]
Thus it suffices to determine the order of $\mathbb E[S_k^2]$.

\medskip
\noindent
Condition on the event $\{D_k=d\}$, where $D_k$ denotes the degree excluding the deterministic self-loop of node $k$. Then
\[
S_k
=
\frac{1}{d+1}
+
\sum_{\ell=1}^{d}\frac{1}{R_\ell+1},
\]
where $R_\ell$ denotes the degree excluding the deterministic self-loop of the $\ell$-th external neighbor of node $k$.

From the explicit preferential attachment neighbor-degree distribution (\cite{fotouhi2013degree}, eq. (2)),
\[
P(r\mid d)
=
\dfrac{m(d+2)}{d\,r(r+1)}
\left[
1-
\dfrac{\binom{2m+2}{m+1}\binom{d+r-2m}{r-m}}
{\binom{d+r+2}{r+1}}
\right],
\]
one obtains
\[
\mathbb E[S_k\mid D_k=d]
=
\beta(m)d+O(1),
\]
where
\[
\beta(m)
=
m\sum_{r=m}^{\infty}\frac{1}{r(r+1)^2}
=
1-m\left(\frac{\pi^2}{6}-H_m^{(2)}\right).
\]
By Jensen's inequality,
\[
\mathbb E[S_k^2\mid D_k=d]
\ge
\left(\mathbb E[S_k\mid D_k=d]\right)^2.
\]
Since $\beta(m)>0$ and
\[
\mathbb E[S_k\mid D_k=d]=\beta(m)d+o(d),
\]
we have
\[
\left(\mathbb E[S_k\mid D_k=d]\right)^2=\Theta(d^2).
\]
Therefore,
\[
\mathbb E[S_k^2\mid D_k=d]\ge \Theta(d^2).
\]
On the other hand,
\[
S_k
\le
1+\frac{d}{m+1},
\]
so
\[
\mathbb E[S_k^2\mid D_k=d]=O(d^2).
\]
Hence
\[
\mathbb E[S_k^2\mid D_k=d]=\Theta(d^2).
\]

\medskip
\noindent
Taking expectations over the degree distribution,
\[
\mathbb E[S_k^2]
=
\sum_{d=m}^{D_{\max}(n)}
\mathbb E[S_k^2\mid D_k=d]\,P_n(D_k=d).
\]
Thus
\[
\mathbb E[S_k^2]
=
\Theta\!\left(
\sum_{d=m}^{D_{\max}(n)} d^2 P_n(D_k=d)
\right).
\]

For preferential attachment,
\[
P_n(D_k=d)
\approx
\frac{2m(m+1)}{d(d+1)(d+2)},
\]
and the maximal degree satisfies
\[
D_{\max}(n)=\Theta(n^{1/2}).
\]
Therefore
\[
\sum_{d=m}^{D_{\max}(n)} d^2 P_n(D_k=d)
=
\Theta\!\left(
\sum_{d=m}^{D_{\max}(n)}\frac{d}{(d+1)(d+2)}
\right).
\]

Using
\[
\frac{d}{(d+1)(d+2)}
=
-\frac{1}{d+1}+\frac{2}{d+2},
\]
the sum reduces to a difference of harmonic sums:
\[
\sum_{d=m}^{D_{\max}(n)}\frac{d}{(d+1)(d+2)}
=
H_{D_{\max}(n)+1}-H_m
+O(1).
\]
Since
\[
H_{D_{\max}(n)+1}
=
\log D_{\max}(n)+O(1)
=
\frac{1}{2}\log n+O(1),
\]
it follows that
\[
\mathbb E[S_k^2]=\Theta(\log n).
\]

Finally,
\[
\mathbb E[Q(n,A)]
=
\frac{1}{n}\mathbb E[S_k^2]
=
\Theta\!\left(\frac{\log n}{n}\right),
\]
which completes the proof.

\end{proof}
% -------------------------------------------------------------

\pagebreak
% =======================================================
\section{Auxiliary Lemmas}
\label{append:proof-aux-lemmas}
% =======================================================

\begin{lemma}
\label{lem:cov-mat-inv}
Let $\bm{\Sigma}$ be a covariance matrix with a common correlation coefficient $\rho \in [0, 1)$ and standard deviations $\sigma_1, \dots, \sigma_n > 0$, such that
\[
\bm{\Sigma}_{ij} = 
\begin{cases}
\sigma_i^2 & \text{if } i = j, \\
\rho\, \sigma_i \sigma_j & \text{if } i \neq j.
\end{cases}
\]
Then, the inverse of $\bm{\Sigma}$ has the following algebraic form
\begin{equation}
\left[\bm{\Sigma}^{-1}\right]_{ij} = 
\begin{cases}
\dfrac{1 + (n-2) \rho}{(1-\rho)\left[1+ (n-1)\rho\right]\sigma_{i}^2} & \text{if } \, i = j, \\
\\
\dfrac{-\rho}{(1-\rho)\left[1+ (n - 1)\rho\right]\sigma_{i}\sigma_{j}}  & \text{if } \, i \neq j.
\end{cases}   
\label{eq:same-diff-sigma}
\end{equation}

In matrix form, letting $\bm{S} = \operatorname{diag}(\sigma_1, \dots, \sigma_n)$ and $\bm{1}$ be the $n$-dimensional column vector of ones, we have:
\begin{equation}
\bm{\Sigma}^{-1} = \bm{S}^{-1} \cdot \left[ \frac{1}{1 - \rho} \left( \bm{I} - \frac{\rho}{1 + (n - 1)\rho} \bm{1}\bm{1}' \right) \right] \cdot \bm{S}^{-1}.
\label{eq:matrix-inverse-general}
\end{equation}

In the special case where $\sigma_i = \sigma$ for all $i$, this expression reduces to
\[
\bm{\Sigma}^{-1} = \frac{1}{\sigma^2(1 - \rho)} \left[ \bm{I} - \frac{\rho}{1 + (n - 1)\rho} \bm{1} \bm{1}' \right].
\]
\end{lemma}

\begin{proof}
The result follows from the Sherman–Morrison formula applied to the equicorrelation structure; see equations (5)–(7) in \cite{clemen1985limits}.
\end{proof}

% ---------------------------------
\begin{lemma}
\label{lem:sumpowerAsumd}
Let $G(n,A)$ be an undirected connected network on $n$ nodes and adjacency matrix $A$. For any integer $k \geq 1$, let $a_{ij}^{(k)}$ denote the $(i,j)$ entry of the matrix power $A^k$. Let $D$ be the diagonal matrix with elements $(D)_{ii} = d_i$. Then,
\begin{equation}
\label{eq:sumpowerAsumd}
\sum_{i,j} a_{ij}^{(k)} \leq \sum_i d_i^k,
\end{equation}
with equality if and only if $G(n,A)$ is regular or $k \leq 2$.
\end{lemma} % ---------------
\begin{proof}
The result follows from \cite{fiol2009number}.
\end{proof}
%-----------------------------------

\begin{lemma}
\label{lem:boundsumsquaresdegrees-loops}
Let $G$ be an undirected network on $n\ge 2$ nodes with self-loops at every node. Let $G^0$ be the simple graph obtained from $G$ by removing self-loops. Denote by $e$ the number of edges in $G^0$, by $d_i^0$ the degree of node $i$ in $G^0$, and by
\[
d_i=d_i^0+1
\]
the degree of node $i$ in $G$, including the self-loop. Then,
\begin{equation}
\label{eq:boundsumsquaresdegrees-loops}
\frac{(2e+n)^2}{n}
\leq
\sum_{i=1}^n d_i^2
\leq
e\left(\frac{2e}{n-1}+n+2\right)+n.
\end{equation}
\end{lemma}

\begin{proof}
Let $G^0$ denote the simple graph obtained after removing the deterministic self-loops. By \cite{de1998upper} bound for simple graphs,
\[
\frac{4e^2}{n}
\leq
\sum_{i=1}^n (d_i^0)^2
\leq
e\left(\frac{2e}{n-1}+n-2\right).
\]
Since the network $G$ contains one self-loop at every node, the degree in $G$ is $d_i=d_i^0+1$.

Therefore,
\[
\sum_{i=1}^n d_i^2
=
\sum_{i=1}^n (d_i^0+1)^2
=
\sum_{i=1}^n (d_i^0)^2
+
2\sum_{i=1}^n d_i^0
+
n.
\]
Because $G^0$ is simple with $e$ edges, $\sum_{i=1}^n d_i^0=2e$. Hence,
\[
\sum_{i=1}^n d_i^2
=
\sum_{i=1}^n (d_i^0)^2+4e+n.
\]
Substituting the lower and upper bounds from \cite{de1998upper} result yields
\[
\frac{4e^2}{n}+4e+n
\leq
\sum_{i=1}^n d_i^2
\leq
e\left(\frac{2e}{n-1}+n-2\right)+4e+n.
\]

Finally, since
\[
\frac{4e^2}{n}+4e+n
=
\frac{(2e+n)^2}{n}
\]
and
\[
e\left(\frac{2e}{n-1}+n-2\right)+4e+n
=
e\left(\frac{2e}{n-1}+n+2\right)+n,
\]
we obtain the equivalent formulation in \eqref{eq:boundsumsquaresdegrees-loops}.
\end{proof}

For the following Lemmas, we define, for each $k\in\{1,\dots,n\}$,
\[
S_k \;=\; \sum_{i=1}^n \frac{A_{ik}}{d_i},
\] and note that
\[
\frac{1}{n}\sum_{k=1}^n S_k = 1
\] holds for every adjacency realization.

% ----------------------------------------
\begin{lemma}
\label{lem:PA-ES-given-d}

Let $D_k$ denote the degree excluding the deterministic self-loop of node $k$ in the preferential attachment model, and let $d_k=D_k+1$ be the corresponding total degree after adding the self-loop. Then, for every $d\ge m$,
\[
\mathbb E[S_k\mid D_k=d]
=
\frac{1}{d+1}
+
d \sum_{r=m}^{\infty}\frac{1}{r+1}\,P(r\mid d),
\]
where $P(r\mid d)$ is the degree distribution of a randomly chosen external neighbor of a node of degree excluding the deterministic self-loop $d$.
\end{lemma}

% -------------------------
\begin{proof}
Fix a node $k$ and condition on the event $\{D_k=d\}$.
Then node $k$ has exactly $d$ external neighbors and one self-loop. Hence
\[
S_k
=
\frac{1}{d+1}
+
\sum_{\ell=1}^{d}\frac{1}{R_\ell+1},
\]
where $R_\ell$ denotes the degree excluding the deterministic self-loop of the $\ell$-th external neighbor of node $k$. Taking conditional expectation,
\[
\mathbb E[S_k\mid D_k=d]
=
\mathbb E\!\left[
\frac{1}{d+1}
+
\sum_{\ell=1}^{d}\frac{1}{R_\ell+1}
\,\Bigg|\, D_k=d
\right].
\]
Since $1/(d+1)$ is deterministic under the conditioning,
\[
\mathbb E[S_k\mid D_k=d]
=
\frac{1}{d+1}
+
\sum_{\ell=1}^{d}
\mathbb E\!\left[\frac{1}{R_\ell+1}\,\Bigg|\,D_k=d\right].
\]
Now each external neighbor of a node of degree $d$ has degree distribution $P(r\mid d)$, so
\[
\mathbb E\!\left[\frac{1}{R_\ell+1}\,\Bigg|\,D_k=d\right]
=
\sum_{r=m}^{\infty}\frac{1}{r+1}\,P(r\mid d).
\]
This expression does not depend on $\ell$, so
\[
\mathbb E[S_k\mid D_k=d]
=
\frac{1}{d+1}
+
d\sum_{r=m}^{\infty}\frac{1}{r+1}\,P(r\mid d).
\]
This proves the result.
\end{proof}

% ----------------------------------------
\begin{lemma}[Asymptotic linearity of $E(S \mid D=d)$]
\label{lem:PA-slope}

For fixed $m\ge 1$,
\[
\mathbb E[S_k\mid D_k=d]
=
\beta(m)\,d+o(d)
\qquad\text{as }d\to\infty,
\]
where
\[
\beta(m)
=
m\sum_{r=m}^{\infty}\frac{1}{r(r+1)^2}
=
1-m\left(\frac{\pi^2}{6}-H_m^{(2)}\right),
\]
and
\[
H_m^{(2)}=\sum_{j=1}^m \frac{1}{j^2}.
\]
\end{lemma}
% ----------------------------------------

\begin{proof}
By Lemma~\ref{lem:PA-ES-given-d},
\[
\mathbb E[S_k\mid D_k=d]
=
\frac{1}{d+1}
+
d \sum_{r=m}^{\infty}\frac{1}{r+1}\,P(r\mid d).
\]
We first study the limit of $P(r\mid d)$ as $d\to\infty$ for fixed $r$.

Using the explicit formula (see \cite{fotouhi2013degree}, eq. (2)),
\[
P(r\mid d)
=
\dfrac{m(d+2)}{d\,r(r+1)}
\left[
1-
\dfrac{\binom{2m+2}{m+1}\binom{d+r-2m}{r-m}}
{\binom{d+r+2}{r+1}}
\right].
\]
As $d\to\infty$,
\[
\frac{d+2}{d}\to 1.
\]
Now consider the combinatorial ratio
\[
\frac{\binom{d+r-2m}{r-m}}{\binom{d+r+2}{r+1}}.
\]
For fixed $m$ and fixed $r$, the numerator is a polynomial in $d$ of degree $r-m$, while the denominator is a polynomial in $d$ of degree $r+1$. Therefore the ratio is of order $d^{-(m+1)}$, and hence tends to zero as $d\to\infty$. It follows that
\[
P(r\mid d)\to \frac{m}{r(r+1)}.
\]

Substituting into the formula for $\mathbb E[S_k\mid D_k=d]$, we obtain
\[
\sum_{r=m}^{\infty}\frac{1}{r+1}P(r\mid d)
\to
\sum_{r=m}^{\infty}\frac{1}{r+1}\cdot \frac{m}{r(r+1)}
=
m\sum_{r=m}^{\infty}\frac{1}{r(r+1)^2}.
\]
Define
\[
\beta(m):=m\sum_{r=m}^{\infty}\frac{1}{r(r+1)^2}.
\]
Hence,
\[
\sum_{r=m}^{\infty}\frac{1}{r+1}P(r\mid d)
=
\beta(m)+o(1).
\]
Substituting back into the expression for $\mathbb E[S_k\mid D_k=d]$ gives
\[
\mathbb E[S_k\mid D_k=d]
=
\frac{1}{d+1}
+
d\,[\beta(m)+o(1)].
\]
Therefore,
\[
\mathbb E[S_k\mid D_k=d]
=
\beta(m)d+o(d).
\]

It remains to simplify $\beta(m)$.
Use the identity
\[
\frac{1}{r(r+1)^2}
=
\frac{1}{r}-\frac{1}{r+1}-\frac{1}{(r+1)^2}.
\]
Indeed,
\[
\frac{1}{r}-\frac{1}{r+1}-\frac{1}{(r+1)^2}
=
\frac{(r+1)^2-r(r+1)-r}{r(r+1)^2}
=
\frac{1}{r(r+1)^2}.
\]
Therefore
\[
\sum_{r=m}^{\infty}\frac{1}{r(r+1)^2}
=
\sum_{r=m}^{\infty}\left(
\frac{1}{r}-\frac{1}{r+1}
\right)
-
\sum_{r=m}^{\infty}\frac{1}{(r+1)^2}.
\]
The first sum telescopes:
\[
\sum_{r=m}^{\infty}\left(
\frac{1}{r}-\frac{1}{r+1}
\right)
=
\frac{1}{m}.
\]
The second sum is
\[
\sum_{r=m}^{\infty}\frac{1}{(r+1)^2}
=
\sum_{s=m+1}^{\infty}\frac{1}{s^2}
=
\frac{\pi^2}{6}-H_m^{(2)}.
\]
Hence
\[
\beta(m)
=
m\left[
\frac{1}{m}-\left(\frac{\pi^2}{6}-H_m^{(2)}\right)
\right]
=
1-m\left(\frac{\pi^2}{6}-H_m^{(2)}\right).
\]
This proves the claim.
\end{proof}

% ----------------------------------------
\begin{lemma}
\label{lem:PA-second-moment-given-d}

For fixed $m\ge 1$,
\[
\mathbb E[S_k^2\mid D_k=d]=\Theta(d^2)
\qquad\text{as }d\to\infty.
\]
More precisely, there exist positive constants $c_1(m),c_2(m)$ such that, for all sufficiently large $d$,
\[
c_1(m)\,d^2
\;\le\;
\mathbb E[S_k^2\mid D_k=d]
\;\le\;
c_2(m)\,d^2.
\]
\end{lemma}
% ----------------------------------------

\begin{proof}
Fix $m\ge 1$.

\medskip
\noindent\textbf{Step 1: lower bound.} By Jensen's inequality, we have
\[
\mathbb E[S_k^2\mid D_k=d]
\;\ge\;
\big(\mathbb E[S_k\mid D_k=d]\big)^2.
\]
By Lemma~\ref{lem:PA-slope},
\[
\mathbb E[S_k\mid D_k=d]
=
\beta(m)\,d+O(1)
\qquad\text{as }d\to\infty,
\]
with $\beta(m)>0$.
Therefore, for all sufficiently large $d$,
\[
\mathbb E[S_k\mid D_k=d]\ge \frac{\beta(m)}{2}\,d.
\]
Hence
\[
\mathbb E[S_k^2\mid D_k=d]
\;\ge\;
\left(\frac{\beta(m)}{2}d\right)^2
=
\frac{\beta(m)^2}{4}\,d^2.
\]
This proves the lower bound with
\[
c_1(m)=\frac{\beta(m)^2}{4}.
\]

\medskip
\noindent\textbf{Step 2: upper bound.} Condition on $\{D_k=d\}$.
Then
\[
S_k
=
\frac{1}{d+1}
+
\sum_{\ell=1}^{d}\frac{1}{R_\ell+1}.
\]
Since every degree excluding the deterministic self-loop satisfies $R_\ell\ge m$, we have
\[
\frac{1}{R_\ell+1}\le \frac{1}{m+1}.
\]
Therefore
\[
S_k
\le
\frac{1}{d+1}
+
d\cdot \frac{1}{m+1}.
\]
Now $\frac{1}{d+1}\le 1$, so
\[
S_k \le 1+\frac{d}{m+1}.
\]
Squaring both sides,
\[
S_k^2
\le
\left(1+\frac{d}{m+1}\right)^2.
\]
Taking conditional expectations,
\[
\mathbb E[S_k^2\mid D_k=d]
\le
\left(1+\frac{d}{m+1}\right)^2.
\]
For all $d\ge 1$,
\[
\left(1+\frac{d}{m+1}\right)^2
\le
2+\frac{2d^2}{(m+1)^2}
\le
\left(2+\frac{2}{(m+1)^2}\right)d^2,
\]
where the last inequality uses $d^2\ge 1$.
Hence, for all sufficiently large $d$,
\[
\mathbb E[S_k^2\mid D_k=d]
\le
c_2(m)\,d^2
\]
with
\[
c_2(m)=2+\frac{2}{(m+1)^2}.
\]

Combining the lower and upper bounds yields
\[
c_1(m)\,d^2
\le
\mathbb E[S_k^2\mid D_k=d]
\le
c_2(m)\,d^2
\]
for all sufficiently large $d$, which proves
\[
\mathbb E[S_k^2\mid D_k=d]=\Theta(d^2).
\]
\end{proof}

%%%%%%%%%%%%%%%%%%%%%%%%%%%%%%%%%%%%%%%%%%%%%%%%%%%%%%%%%%%%%%%%%%%
\end{document}